\documentclass[a4paper,11pt]{article}

\pdfoutput=1 % if your are submitting a pdflatex (i.e. if you have
\usepackage{jheppub} % for details on the use of the package, please
\usepackage{nicefrac}
\usepackage[T1]{fontenc} % if needed

\usepackage{graphicx}

\usepackage{MnSymbol}
\usepackage{empheq}
\usepackage{slashed} %For slash notation  

\usepackage{tikz}
\usetikzlibrary{decorations}

\title{\boldmath Massive Spinning Bosons on the Celestial Sphere}

\author{Y. T. Albert Law,$^1$}
\author{Michael Zlotnikov,$^1$}

\affiliation{$^1$ Department of Physics, Center for Theoretical Physics,
Columbia University, 538 West 120th Street, New York, NY 10027, USA.}
\emailAdd{yal2109@columbia.edu}
\emailAdd{mz2737@columbia.edu}

\abstract{\\
A natural extension of the Pasterski-Shao-Strominger (PSS) prescription is described, enabling the map of Minkowski space amplitudes with massive spinning external legs to the celestial sphere to be performed. An integral representation for the conformal primary wave function (CPW) of massive spinning bosons on the celestial sphere is derived explicitly for spin-one and -two. By analogy with the spin-zero case, the spinning bulk-to-boundary propagator on Euclidean AdS is employed to extend the massive CPW integral representation to arbitrary integer spin, and to describe the appropriate inverse transform of massive spinning CPWs back to the plane wave basis in Minkowski space. Subsequently, a massive spin-$s$ momentum operator representation on the celestial sphere is determined, and used in conjunction with known Lorentz generators to derive Poincar\'e symmetry constraints on generic massive spinning two-, three- and four-point celestial amplitude structures. Finally, as a consistency check, three-point Minkowski space amplitudes of two massless scalars and a spin-one or -two massive boson are explicitly mapped to the celestial sphere, and the resulting three-point function coefficients are confirmed to be in exact agreement with the results obtained from Poincar\'e symmetry constraints.
}

\begin{document} 
\maketitle
\flushbottom

\section{Introduction}

Minkowski space scattering amplitudes can be equivalently mapped to the celestial sphere at light-like infinity, where they are encoded in terms of conformal correlator objects called celestial amplitudes. This map for amplitudes of massive scalars and massless particles of various spin was described by Pasterski, Shao and Strominger (PSS) \cite{Pasterski:2016qvg,Pasterski:2017kqt}. Ideally, the goal is to discover so called Celestial Conformal Field Theories (CCFT) that would give rise to celestial amplitudes for specific scattering processes of interest, entirely from the celestial sphere perspective without reference to Minkowski space amplitudes. While to this date no example of a CCFT is known, note that since the map to the celestial sphere is bijective, celestial amplitudes can also be considered more pragmatically as a new basis for the original Minkowski space amplitudes. Alternative descriptions of scattering amplitudes can potentially reveal hidden features that were not previously accessible from other points of view. Considering that asymptotic states that serve as input to scattering amplitudes originate at or near the celestial sphere, investigation of an amplitude description entirely on this boundary without reference to the bulk of Minkowski space is particularly interesting.

Celestial amplitudes and their properties are actively being studied: The residues of tree-level Minkowski space amplitudes were mapped to the celestial sphere in \cite{Cardona:2017keg}. Explicit amplitude examples were mapped to the celestial sphere \cite{Pasterski:2016qvg,Lam:2017ofc,Banerjee:2017jeg,Nandan:2019jas,Pasterski:2017ylz,Schreiber:2017jsr,Stieberger:2018edy,Puhm:2019zbl,Guevara:2019ypd},
and an alternative map was proposed and studied in \cite{Banerjee:2018gce,Banerjee:2018fgd,Banerjee:2019prz}.
Conformal soft theorems for celestial amplitudes were derived in \cite{Donnay:2018neh,Banerjee:2019aoy,Fan:2019emx,Pate:2019mfs,Nandan:2019jas,Adamo:2019ipt,Puhm:2019zbl,Guevara:2019ypd,Himwich:2019dug}. 
In \cite{Lam:2017ofc,Nandan:2019jas} conformal partial wave decomposition of celestial amplitudes was discussed.
 BMS symmetry realization on the celestial sphere, and OPE expansions of celestial operators appeared in \cite{Ball:2019atb,Fotopoulos:2019tpe,Pate:2019lpp,Banerjee:2020kaa,Fotopoulos:2019vac}.

In \cite{Law:2019glh} we have derived the explicit constraints on celestial amplitudes implied by Poincar\'e symmetry (based on conformal Ward identities and momentum conservation), making use of the representation of Lorentz and massless momentum generators on the celestial sphere \cite{Stieberger:2018onx}. Considering that conformal primary wave functions and the map to the celestial sphere for amplitudes involving spinning massive particles have not been discussed in the literature to date, the constraints presented in \cite{Law:2019glh} omit the case of spinning massive particles. In this work we set out to fill in these gaps, and obtain the following results.

In the massive case, we take the Ansatz that the plane wave basis functions of an entire spin multiplet are to be mapped to the celestial sphere collectively, instead of a single spin component value in the direction of propagation. This is well motivated by the fact that a single spin value in direction of propagation is not a conserved quantity under Poincar\'e transformations, while on the other hand the entire multiplet transforms into itself. Demanding the appropriate transformation properties on the celestial sphere, we determine the explicit integral representation for spin-1 and -2 massive conformal primary wave functions (CPW) by solving differential equations derived from a Lorentz transformation infinitesimally close to the identity. The integral representation is in line with the massive scalar case discussed in \cite{Pasterski:2016qvg}. By analogy with the scalar case, we point out that the integration weight in the integral representation of the massive conformal primary wave function of arbitrary spin is given by a respective bulk-to-boundary propagator on Euclidean Anti-de Sitter space $H_3$. This relation also reveals a straightforward inverse transform of the massive conformal primary wave function back to Minkowski space, based on an integral over the split representation of a harmonic function known in the literature \cite{Costa:2014kfa}.

A natural extension of Pasterski-Shao-Strominger (PSS) prescription, to enable the map of Minkowski space amplitudes with massive spinning external legs to the celestial sphere, follows directly from the conformal primary wave function discussion by recalling that amplitudes are multi-linear in polarization tensors, such that the polarization part of the integration weight for the integral transform hides within the amplitudes.

With the integral representation of the massive conformal primary wave function in hand, we derive a corresponding massive momentum operator representation on the celestial sphere, which has the appropriate momentum eigenvalue in the bulk. The resulting momentum operator turns out to be spin dependent and non-diagonal in spin multiplet space. As required, this momentum operator properly closes the Poincar\'e algebra, which proves that it is indeed correct for arbitrary integer spin $s$.

Making use of this newly discovered massive spinning momentum operator representation, we repeat the analysis of \cite{Law:2019glh} and derive the explicit constraints on two-, three- and four-point structures implied by conformal Ward identities and momentum conservation in the case when massive spinning external particles are present. 

In the two-point case we find that the resulting recurrence relations for the two-point function coefficient can be solved. However, as in the massive scalar case \cite{Law:2019glh}, the Mellin-like integral in the inverse transform to Minkowski space diverges, such that there is no corresponding two-point amplitude counterpart in Minkowski space.

In the three-point case with two massless and one massive particle we show how the three-point function coefficient for the entire spin multiplet of the massive particle can be built up iteratively. The expected property that the difference of helicity values of the two massless particles produced in the decay of a massive particle may not exceed the maximum spin that can be accommodated by the spin multiplet of the massive particle, emerges as a multiplet truncation condition from the celestial sphere perspective.

Generically, the momentum conservation constraints admit three-point function coefficient solutions that allow for periodic dependency on individual conformal dimensions $\Delta=1+i\lambda$ of the external particles. As we discussed in detail at the end of section 4 in \cite{Law:2019glh}, periodic holomorphic functions of finite order are unbounded on any complex vertical line, which spoils the convergence of the Mellin-like integration over $-\infty<\lambda<\infty$ in the inverse transform back to Minkowski space. To ensure convergence, the three-point function coefficient better have no non-trivial periodicities in $\Delta_i$.

In three-point cases with two massive and one massless particle, or three massive particles, as well as the four-point case with one massive and three massless particles, we report the explicit resulting recurrence and differential equations for three-point function coefficient and four-point function of conformal cross ratios. While these equations are hard to solve in general, they are likely to be useful as consistency requirements when working with specific amplitude examples on the celestial sphere.

Finally, as a cross check, we perform the map of tree-level amplitudes of two massless scalars and one massive spin-1 or -2 particle from Minkowski space to the celestial sphere. We find the expected three-point function coordinate dependence, and the resulting three-point function coefficients perfectly agree with our results obtained purely from symmetry.

This work is organized as follows. In section \ref{sec:CPWspin} we recall the problem setting, introduce spinning massive conformal primary wave functions and demonstrate their inverse transform back to the plane wave basis in Minkowski space. In section \ref{sec:smP} we present the spinning massive momentum operator representation on the celestial sphere. Section \ref{sec:234constr} investigates the explicit constraints implied by conformal Ward identities and momentum conservation on two-, three- and four-point structures involving massive spinning particles. Appendix \ref{s1s2polarizations} parametrizes massive spin-1 and -2 polarization tensors, appendix \ref{op} discusses the formalism of encoding symmetric transverse traceless tensors in terms of polynomials, while appendix \ref{sec:3pexamples} shows examples of three-point Minkowski space amplitudes mapped to the celestial sphere.

\section{Spinning massive conformal primary wave functions}
\label{sec:CPWspin}
\subsection{On-shell massive and massless momentum parametrization in the bulk}
As described in \cite{Pasterski:2016qvg}, the mass-shell of a massive momentum vector is parametrized by a hyperbolic slice of Minkowski space. This $H_3$ slice in Poincar\'e coordinates $y,z,\bar z$ has the metric
\begin{align}
ds_{H_3}^2=\frac{dy^2+dz d\bar z}{y^2}~~~\text{with}~~~0<y<\infty\,,\text{ and}~z,\bar z\in\mathbb{C}\,,
\end{align}
and features the $SL(2,\mathbb{C})$ isometry
\begin{align}
z\to \frac{(a z+b)(\bar{c}\bar{z}+\bar{d})+a\bar{c} y^2}{(c z+d)(\bar{c} \bar{z}+\bar{d})+c \bar{c} y^2}~,~~\bar{z}\to \frac{(\bar{a} \bar{z}+\bar{b})(c  z+  d)+\bar{a}  c y^2}{(\bar{c}\bar{z}+\bar{d})( c z+d)+c \bar{c}y^2}~,~~y\to \frac{y}{(c z+d)(\bar{c}\bar{z}+\bar{d})+c\bar{c} y^2},
\end{align}
where $\bar a=a^*,\bar b=b^*,\bar c=c^*,\bar d=d^*,\in\mathbb{C}$ and $ad-bc=\bar a\bar d-\bar b\bar c=1$. Such an $H_3$ slice asymptotes to the so called celestial sphere at the boundary $y\to 0$. The $SL(2,\mathbb{C})$ transformations on $H_3$ entail M\"obius transformations of complex coordinates $w,\,\bar w$ on the celestial sphere, which involve the same transformation parameters:
\begin{align}
\label{moeb}
w\to\frac{a w+b}{c w+d}~~,~~~\bar{w}\to\frac{\bar{a} \bar{w}+\bar{b}}{\bar{c} \bar{w}+\bar{d}}.
\end{align}
The embedding of a mass $m$ on-shell ($(m\hat p)^2=-m^2$) momentum into Minkowski space $m\hat{p}^{\mu}:H_3\to\mathbb{R}^{1,3}$ is then
\begin{align}
\label{mphat}
m \hat p^{\mu}(y,z,\bar{z})=m\left(\frac{1+y^2+z\bar{z}}{2y},\frac{\bar{z}+z}{2y},i\frac{\bar{z}-z}{2y},\frac{1-y^2-z\bar{z}}{2y}\right),
\end{align}
while a massless (light-like $(\omega q)^2=0$) momentum pointing to a point on the celestial sphere amounts to the vector
\begin{align}
\label{omegaq}
\omega q^{\mu}(w,\bar w)&=\omega (1+w \bar{w},~\bar{w} + w,~i(\bar{w}-w),~1-w \bar{w}),
\end{align}
with an energy scale $\omega$. This parametrization satisfies the constraint
\begin{align}
\label{vDef}
\frac{1}{2}q^+=q^\mu v_\mu =1\quad\text{with}\quad v^\mu=-\frac{1}{2}\partial_w\partial_{\bar w}q^\mu=(-\nicefrac{1}{2},0,0,\nicefrac{1}{2}).
\end{align}
We work with a projection of the celestial sphere onto a complex plane $\mathbb{C}$, which has the metric\footnote{Note that the parametrization of $q^\mu$ actually produces an induced metric of a complex plane on the boundary. Equivalently, we could work with an induced metric of a complex sphere instead, if we redefine $\omega\to\frac{\omega'}{1+w\bar w}$, such that $q^\mu$ effectively receives an extra factor $\frac{1}{1+w\bar w}$.}
\begin{align}
ds^2 = 4 dw d\bar{w}\quad \text{or} \quad g_{w \bar{w}}=2,
\end{align}
while the inverse bulk metric can be expressed as
\begin{align}\label{d=2 metric}
\eta^{\mu\nu}=\frac{1}{2}\left(\frac{\partial q^\mu}{\partial w}\frac{\partial q^\nu}{\partial \bar{w}}+\frac{\partial q^\mu}{\partial \bar{w}}\frac{\partial q^\nu}{\partial w}\right)+q^\mu v^\nu +v^\mu q^\nu.
\end{align}

The explicit Lorentz transformation matrix $\Lambda^\mu{}_\nu$, that acts on four-vectors in the embedding Minkowski space as
\begin{align}
m\hat p^\mu\to m\hat p'^\mu= m\Lambda^\mu{}_\nu\hat p^\nu~,~~~\omega q^\mu\to\omega'q'^\mu= \omega\Lambda^\mu{}_\nu q^\nu~,~~~\text{with}~~~\omega\to\omega'=|cw+d|^2\omega\,,
\end{align}
 and corresponds to the M\"obius transformation (\ref{moeb}) on the celestial sphere, is given by
\begin{align}
\label{lorentzL}
\Lambda^\mu{}_\nu=\frac{1}{2}\left(
\begin{array}{cccc}
 a \bar{a}+b \bar{b}+c \bar{c}+d \bar{d} & b \bar{a}+a \bar{b}+d \bar{c}+c \bar{d} & i
   \left(-b \bar{a}+a \bar{b}-d \bar{c}+c \bar{d}\right) & -a \bar{a}+b \bar{b}-c
   \bar{c}+d \bar{d} \\
 c \bar{a}+a \bar{c}+d \bar{b}+b \bar{d} & d \bar{a}+a \bar{d}+c \bar{b}+b \bar{c} & i
   \left(-d \bar{a}+a \bar{d}+c \bar{b}-b \bar{c}\right) & -c \bar{a}-a \bar{c}+d
   \bar{b}+b \bar{d} \\
 i \left(c \bar{a}-a \bar{c}+d \bar{b}-b \bar{d}\right) & i \left(d \bar{a}-a
   \bar{d}+c \bar{b}-b \bar{c}\right) & d \bar{a}+a \bar{d}-c \bar{b}-b \bar{c} & i
   \left(-c \bar{a}+a \bar{c}+d \bar{b}-b \bar{d}\right) \\
 -a \bar{a}-b \bar{b}+c \bar{c}+d \bar{d} & -b \bar{a}-a \bar{b}+d \bar{c}+c \bar{d} &
   i \left(b \bar{a}-a \bar{b}-d \bar{c}+c \bar{d}\right) & a \bar{a}-b \bar{b}-c
   \bar{c}+d \bar{d} \\
\end{array}
\right),
\end{align}
while the inverse $\Lambda^{-1}{}^\mu{}_\nu$ is obtained by exchanging $a\leftrightarrow d,\,b\to-b,\,c\to-c$, and the same for the barred parameters.

\subsection{Map of spinning massive amplitudes to the celestial sphere}
The map of Minkowski space amplitudes $\mathcal{A}_n$ with massless external particles of various spin to celestial amplitudes $A_n$ on the celestial sphere at light-like infinity has been derived in \cite{Pasterski:2017kqt} to be given by the Mellin transform in the energy scale $\omega_j$ of the $j$-th massless external particle momenta
\begin{align}
\label{masslessAmap}
{A_n}=...\left(\prod_{j}\int_0^\infty d\omega_j\,\omega_j^{i\lambda}\right)\mathcal{A}_n,
\end{align}
in an appropriate polarization gauge. Here, the leading ellipsis denote suppressed transforms for potentially present massive particles, to be discussed below. Parameters $\lambda_j$ enter the conformal operator dimensions 
\begin{align}
\Delta_j=1+i\lambda_j~~~\text{with}~~~\lambda\in\mathbb{R}
\end{align}
in continuous series representation, so that particle properties are encoded by operator insertions on the celestial sphere.
Since massless particles reach specific points at light-like infinity directly, their helicity values $l_j$ map diagonally to the spin values $J_j$ of corresponding operator insertions.

For the case of external massive scalar particles in a Minkowski space amplitude, the map to celestial amplitudes was obtained in \cite{Pasterski:2016qvg}, for each $j$-th massive external scalar, to be given by
\begin{align}
\label{scalarAmap}
A_n&=...\left(\prod_j\int_0^\infty\frac{dy_j}{y_j^3}\int dz_j d\bar z_j \,(-q_j\cdot \hat p_j)^{-\Delta_j}\right) \mathcal{A}_n,
\end{align}
where the leading ellipsis denote suppressed transformations for other types of external particles in $\mathcal{A}_n$. The integral transformation weight $(-q_j\cdot \hat p_j)^{-\Delta}$ is the so called scalar bulk to boundary propagator on $H_3$ \cite{Costa:2014kfa}. The integral transform collects contributions from the entire mass shell $y_j,z_j,\bar z_j$ and returns a localized insertion point $w_j,\bar w_j$ on the celestial sphere for $j$-th external scalar. It is our objective to generalize this mapping prescription to accommodate the cases of massive spinning external particles.

Recall that amplitudes in Minkowski space are multi-linear in polarization tensors of external particles
\begin{align}
{\mathcal{A}_n}_{b_1,...,b_j}=\left(\prod_{i=1}^j\epsilon_{b_i}^{\mu_{1_i}...\mu_{s_i}}\right)\mathcal{A}_n{}_{\mu_{1_1}...\mu_{s_1},...,\mu_{1_j}...\mu_{s_j}},
\end{align}
with as many sets of $s_j$ Lorentz indices on the uncontracted amplitude as there are massive spinning particles with polarization tensors $\epsilon_{b_i}$ involved in the process.\footnote{See appendix \ref{s1s2polarizations} for explicit examples of spin $s=1$ and $s=2$ massive polarizations.} The subscripts $b_i$ are spin (in direction of propagation) multiplet indices for the massive particles.

Considering that a Lorentz transformation in the bulk can flip the spin component along the axis of propagation of a massive particle, it is natural to propose that an amplitude of definite spin particles on the celestial sphere is to be obtained from a combination of all spin component values of the respective spin multiplets in the bulk. Therefore, we seek a map where each massive external leg of an amplitude is transformed to the celestial sphere as
\begin{align}
\label{massAmap}
{A_n}_{J_1,...,J_j}=...\left(\prod_{i=1}^j \int \frac{dy_i}{y^3_i}dz_i d\bar z_i \sum_{b_i=-s_i}^{s_i}G^{(s_i)}_{J_i b_i}\right){\mathcal{A}_n}_{b_1,...,b_j},
\end{align}
 with a spin $s_i$ integration weight matrix $G^{(s_i)}_{J_ib_i}$ for massive particles. Here, $J_i$ are spin indices on the celestial sphere $J_i=-s_i,-s_i+1,...,s_i$. The leading ellipsis denote suppressed integral transforms for possibly present massless particles (\ref{masslessAmap}), which are unchanged. 

In the next subsection we describe how the integration weight matrix $G^{(s_i)}_{J_ib_i}$ can be obtained from the consideration of spinning massive conformal primary wave functions.

\subsection{Spin-zero, -one and -two massive conformal primary wave functions}
As defined in \cite{Pasterski:2016qvg}, the conformal primary wavefunction of a mass $m$ scalar (outgoing/incoming: $\pm$) on the celestial sphere can be obtained via the integral transform of the plain wave basis in Minkowski space
\begin{align}
\label{s0Phi}
\phi_{\pm,\Delta,m}(X;w,\bar w)&=\int_0^\infty\frac{dy}{y^3}\int dz d\bar z \,\left(-q\cdot \hat p\right)^{-\Delta} e^{\pm i m\hat p^\nu X_\nu},
\end{align}
analogously to (\ref{scalarAmap}). In this section we set up an ad-hoc procedure to obtain the mapping of the massive spin $s$ plain wave basis
\begin{align}
{\phi^{\mu_1...\mu_s}_{\pm,m,_b}}(X)=\epsilon_b^{\mu_1...\mu_s}e^{\pm i m\hat p^\nu X_\nu},
\end{align}
with polarization tensor $\epsilon_b^{\mu_1...\mu_s}$ for spin component value $b=-s,-s+1,...,s$ in direction of propagation, to the celestial sphere. 

Guided by our expectation of how amplitudes involving spinning massive particles are to be mapped (\ref{massAmap}), we seek an integration weight matrix $G^{(s)}_{J b}$ for the same integral transform over the hyperbolic on-shell slice providing the massive spinning conformal primary wave function
\begin{align}
\label{phiAnyS}
{\phi^{\mu_1...\mu_s}_{\pm,\Delta,m}}_J(X;w,\bar w)&=\int_0^\infty\frac{dy}{y^3}\int dz d\bar z \,\sum_{b=-s}^sG^{(s)}_{J b}(w,\bar w;y,z,\bar z)\epsilon_b^{\mu_1...\mu_s}e^{\pm i m\hat p^\nu X_\nu}.
\end{align}

In \cite{Pasterski:2017kqt}, conformal primary wave functions $\phi$ were defined as solutions to the massive Klein-Gordon equation with appropriate covariant transformation properties. The ansatz (\ref{phiAnyS}) trivially satisfies the Klein-Gordon equation, since all plane waves do. Here we exploit the fact that the integration weight matrices $G^{(s)}_{J b}$ are uniquely fixed (up to row normalization) by demanding appropriate covariant transformation properties, which fixes the integral representation of $\phi$.

The desired conformal primary wavefunction $\phi$ must transform as a conformal primary on the celestial sphere, and maintain the usual Lorentz transformation $\Lambda^\mu{}_\nu$ on the Minkowski vector indices. The exponential and the integration measure in (\ref{phiAnyS}) are invariant under such transformation, so that the combination of integration weight matrix and polarization tensors must transform as
\begin{align}
\label{Gtraf}
\begin{split}
\sum_bG^{(s)}_{J b}\epsilon_b^{\mu_1...\mu_s}&\to \sum_b{G'}^{(s)}_{J b}{\epsilon'}_b^{\mu_1...\mu_s}\\
&=(c w+d)^{\Delta+J}(\bar c \bar w+\bar d)^{\Delta-J}\Lambda^{\mu_1}{}_{\nu_1}...\Lambda^{\mu_s}{}_{\nu_s}\sum_bG^{(s)}_{J b}\epsilon_b^{\nu_1...\nu_s}.
\end{split}
\end{align}
Requiring this transformation property, we may consider a Lorentz transformation (\ref{lorentzL}) that is infinitesimally close to the identity:
\begin{align}
a=\frac{1+\alpha \beta}{1-\gamma}~~~,~~~b=\alpha~~~,~~~c=\beta~~~,~~~d=1-\gamma\,,
\end{align}
with $\alpha,\beta,\gamma$ infinitesimal, and similarly for $\bar a,\bar b,\bar c,\bar d$ and $\bar \alpha,\bar\beta,\bar\gamma$. Taking as a convenient ansatz $G^{(s)}_{J b}=(-\hat p\cdot q)^{-\Delta-s}g^{(s)}_{J b}$, using known expressions for the polarization tensors (e.g. (\ref{s1pols}) and (\ref{s2pols}) for spin-one and -two respectively) and expanding (\ref{Gtraf}) to linear order in the infinitesimal transformation parameters, yields a set of coupled first order differential equations in $w,\bar w,z,\bar z,y$ variables for the matrix components $g^{(s)}_{J b}$, which can be easily solved iteratively.

In cases of spin-zero, -one and -two this procedure straightforwardly gives
\begin{align}
\label{Gs0}
&G^{(0)}_{Ja}=(-q\cdot\hat p)^{-\Delta}\,,\\
\label{Gs1}
&G^{(1)}_{Ja}=(-q\cdot\hat p)^{-\Delta-1}\left(
\begin{array}{ccc}
 \frac{\sqrt{2} (w-z)^2}{y} &- 2 (w-z) & -\sqrt{2} y \\
 \sqrt{2} (w-z) & \frac{(w-z) \left(\bar{w}-\bar{z}\right)}{y}-y &
   \sqrt{2} \left(\bar{w}-\bar{z}\right) \\
 \sqrt{2} y & 2 \left(\bar{w}-\bar{z}\right) & -\frac{\sqrt{2}
   \left(\bar{w}-\bar{z}\right)^2}{y} \\
\end{array}
\right)\,,\\
\label{Gs2}
&G^{(2)}_{Ja}=(-q\cdot\hat p)^{-\Delta-2}\cdot\\
&
\resizebox{\textwidth}{!}{$
\cdot\left(
\begin{array}{ccccc}
 \frac{2 (w-z)^4}{y^2} & \frac{4 i (w-z)^3}{y} & 2 \sqrt{6} (w-z)^2 & 4 i y
   (w-z) & 2 y^2 \\
 \frac{2 (w-z)^3}{y} & i (w-z)^2 \left(3-\frac{(w-z)
   \left(\bar{w}-\bar{z}\right)}{y^2}\right) & \frac{\sqrt{6} (w-z)
   \left(y^2-(w-z) \left(\bar{w}-\bar{z}\right)\right)}{y} & i \left(y^2-3 (w-z)
   \left(\bar{w}-\bar{z}\right)\right) & 2 y \left(\bar{z}-\bar{w}\right) \\
 2 (w-z)^2 & \frac{2 i (w-z) \left(y^2-(w-z)
   \left(\bar{w}-\bar{z}\right)\right)}{y} & \frac{\sqrt{\frac{2}{3}}
   \left((w-z) \left(\bar{w}-\bar{z}\right) \left((w-z)
   \left(\bar{w}-\bar{z}\right)-4 y^2\right)+y^4\right)}{y^2} & \frac{2 i
   \left(\bar{z}-\bar{w}\right) \left(y^2-(w-z)
   \left(\bar{w}-\bar{z}\right)\right)}{y} & 2 \left(\bar{z}-\bar{w}\right)^2 \\
 2 y (w-z) & i \left(y^2-3 (w-z) \left(\bar{w}-\bar{z}\right)\right) &
   \frac{\sqrt{6} \left(\bar{z}-\bar{w}\right) \left(y^2-(w-z)
   \left(\bar{w}-\bar{z}\right)\right)}{y} & i \left(\bar{z}-\bar{w}\right)^2
   \left(3-\frac{(w-z) \left(\bar{w}-\bar{z}\right)}{y^2}\right) & \frac{2
   \left(\bar{z}-\bar{w}\right)^3}{y} \\
 2 y^2 & 4 i y \left(\bar{z}-\bar{w}\right) & 2 \sqrt{6}
   \left(\bar{z}-\bar{w}\right)^2 & \frac{4 i
   \left(\bar{z}-\bar{w}\right)^3}{y} & \frac{2
   \left(\bar{z}-\bar{w}\right)^4}{y^2} \\
\end{array}
\right)
	$},\notag
\end{align}
where the $(2s+1)\times(2s+1)$ matrix components are labeled by indices $J,a\in\{-s,...,s\}$, and where we have chosen a particular normalization for each row that will be convenient for the parametrization of the massive momentum operator representation discussed in the following section.

The integration weight matrices above can be written more succinctly by introducing the following building blocks
\begin{align}\label{building blocks}
I^{\mu\nu}\equiv(q\cdot\hat p )\eta^{\mu\nu}-q^\mu \hat p^\nu~~,~~
T_{-1}^\mu \equiv I^{\mu\nu}(\partial_{\bar w}q_\nu)~~,~~T_{0}^\mu \equiv I^{\mu\nu}\hat p_\nu~~,~~T_{+1}^\mu \equiv I^{\mu\nu}(-\partial_{ w}q_\nu)\,.
\end{align}
With this, the spin-1 integration weight matrix is given by
\begin{align}
G^{(1)}_{Ja}=(-q\cdot\hat p)^{-\Delta-1}T_{J}^\mu{\epsilon^*_a}_\mu\,,
\end{align}
while the spin-2 integration weight matrix amounts to
\begin{align}
G^{(2)}_{-2,a}&=(-q\cdot\hat p)^{-\Delta-2}T_{-1}^\mu T_{-1}^\nu{\epsilon^*_a}_{\mu\nu}~~~,~~~G^{(2)}_{-1,a}=(-q\cdot\hat p)^{-\Delta-2}\frac{T_{-1}^\mu T_{0}^\nu+T_{0}^\mu T_{-1}^\nu}{2}{\epsilon^*_a}_{\mu\nu}\,,\notag\\
G^{(2)}_{0,a}&=(-q\cdot\hat p)^{-\Delta-2}\left(\frac{2}{3}T_{0}^\mu T_{0}^\nu+\frac{1}{3}\frac{T_{-1}^\mu T_{+1}^\nu+T_{+1}^\mu T_{-1}^\nu}{2}\right){\epsilon^*_a}_{\mu\nu}\,,\\
G^{(2)}_{+2,a}&=(-q\cdot\hat p)^{-\Delta-2}T_{+1}^\mu T_{+1}^\nu{\epsilon^*_a}_{\mu\nu}~~~,~~~G^{(2)}_{+1,a}=(-q\cdot\hat p)^{-\Delta-2}\frac{T_{+1}^\mu T_{0}^\nu+T_{0}^\mu T_{+1}^\nu}{2}{\epsilon^*_a}_{\mu\nu}\,.\notag
\end{align}
Note that  $\sum_a G^{(s)}_{Ja}\epsilon_a^{\mu...\nu}$ conveniently simplifies and is equal to the above expressions with the complex conjugate polarization factors ${\epsilon_a^*}_{\mu_1...\mu_s}$ dropped.

\subsection{Arbitrary integer spin massive conformal primary wave functions}
While the procedure in the previous section is useful to develop conceptual understanding and investigate a few explicit examples of lower spin, it is too cumbersome to parametrize the arbitrary spin situation. Luckily, by analogy with the fact that the scalar integration weight $(-q\cdot \hat p)^{-\Delta}$ in (\ref{s0Phi}) is the scalar bulk-to-boundary propagator on $H_3$, the product of integration weight matrices with polarization tensors $\sum_a G^{(s)}_{Ja}\epsilon_a^{\mu...\nu}$ is also related to spinning bulk-to-boundary propagators on $H_3$. More precisely, we start with spin-$|J|$ bulk-to-boundary propagator \cite{Costa:2014kfa} (dropping a normalization factor):
\begin{align}\label{bulk to bdy}
\Pi_{\Delta,|J|}(\hat{p},Y; q, Z)=\frac{((\hat{p}\cdot q)(Y\cdot Z)-(\hat{p}\cdot Z)(q\cdot Y))^{|J|}}{(-\hat{p}\cdot q)^{\Delta+|J|}}=\frac{(Y\cdot I \cdot Z)^{|J|}}{(-\hat{p}\cdot q)^{\Delta+|J|}},
\end{align}
where $I_{\mu\nu}$ is one of the building blocks in \eqref{building blocks}. $Y^\mu$ and $Z^\mu$ are auxiliary variables encoding tensor structures on the $H_3$ ($\hat{p}^2+1=0$) slice and the light cone ($q^2=0$) respectively, which are discussed in appendix \ref{op}. Since $\hat{p}^\mu I_{\mu\nu}=0=I_{\mu\nu} q^\nu$, expression \eqref{bulk to bdy} satisfies transversality on both sets of indices
\begin{align}
\hat{p}\cdot \partial_Y \Pi_{\Delta,J}(\hat{p},Y; q, Z)=0=q \cdot \partial_Z \Pi_{\Delta,J}(\hat{p},Y; q, Z).
\end{align}
To obtain an object that transforms as a $\mathbb{R}^{1,3}$ spin-$s$ Lorentz tensor, we take the $(s-|J|)$-fold formal derivative $Y\cdot\nabla_{\hat{p}}=Y\cdot \partial_{\hat{p}}$ :
\begin{align}
\frac{1}{(\Delta+|J|)_{s-|J|}} (Y\cdot \partial_{\hat{p}})^{s-|J|}\Pi_{\Delta,|J|}(\hat{p},Y; q, Z),
\end{align}
where the prefactor is inserted for later convenience, and $(a)_n = \frac{\Gamma(a+n)}{\Gamma(a)}$ is the Pochhammer symbol. Subsequently, as outlined in appendix \ref{lc op}, we use the operator $R_\mu$ defined in (\ref{Rop}) to recover the symmetric traceless transverse tensor encoded by the polynomial in auxiliary $Z^\mu$ variables. The action of $(Dq\cdot R)^{|J|}$ facilitates a contraction of $Dq^\mu$ vectors with the symmetric traceless transverse tensor indices, which (up to normalization) effectively reduces to replacing each $Z^\mu$ by $Dq^\mu$, as outlined in (\ref{DqR1}), (\ref{DqR2}). Finally, we take the plus or minus spin component pull-backs to the celestial sphere with a convenient coefficient, which leads to the definition
\begin{align}
G^{(s)}_{J,\Delta}(\hat{p},Y; w,\bar{w})\equiv&\frac{(-1)^{|J|\theta_{-J}}}{|J|!\left(\frac{d-2}{2}\right)_{|J|}(\Delta+|J|)_{s-|J|}}\left(Dq \cdot R \right)^{|J|} (Y\cdot \partial_{\hat{p}})^{s-|J|} \Pi_{\Delta,|J|}(\hat{p},Y; q, Z)\nonumber\\
\label{GtransSym}
=&(-1)^{|J|\theta_{-J}} \frac{(Y\cdot q)^{s-|J|}((\hat{p}\cdot q)(Y\cdot Dq)-(\hat{p}\cdot Dq)(q\cdot Y))^{|J|}}{(-\hat{p}\cdot q)^{\Delta+s}} ,
\end{align}
where $\theta_x=\theta(x)$ is the Heaviside step function, and
\begin{align}
\label{JDq}
Dq^\mu=\left\{
\begin{array}{ccc}
 \partial_{\bar w}q^\mu&\text{for}&J>0 \\
 \partial_{w}q^\mu&\text{for}&J<0 \\
\end{array}
\right.\,.
\end{align}
As discussed in appendix \ref{op}, a transverse symmetric and traceless tensor with $s$ Minkowski indices is then recovered by applying the special differential operator \eqref{hyper recover op} $s$ times in succession. This directly produces the product of integration weight matrix with polarization tensors, discussed in the previous sub-section\footnote{Due to the orthogonality of polarization tensors, we then recover $G^{(s)}_{Ja}=\left(\sum_b G^{(s)}_{Jb}\epsilon_b^{\mu_1...\mu_s}\right){\epsilon_a^*}_{\mu_1...\mu_s}$.} 
\begin{align}
\label{GEanyS}
\sum_b G^{(s)}_{Jb}\epsilon_b^{\mu_1...\mu_s}=\frac{1}{s!(\frac{1}{2})_s}K^{\mu_1}...K^{\mu_s} G^{(s)}_{J,\Delta}(\hat{p},Y; w,\bar{w}).
\end{align}
It is trivial to verify, that the explicit examples $\sum_b G^{(1)}_{Jb}\epsilon_b^{\mu}$ and $\sum_b G^{(2)}_{Jb}\epsilon_b^{\mu\nu}$ we have considered in more detail are readily recovered in this fashion.

Using expression (\ref{GtransSym}) we can define a polynomial encoding of the integral representation of massive spinning conformal primary wave function with arbitrary spin
\begin{align}
\phi^{\pm,m,s}_{J,\Delta}(X,{Y} ;w,\bar{w})=\int_0^\infty\frac{dy}{y^3}\int dz d\bar z \,G^{(s)}_{J,\Delta}(\hat{p},Y; w,\bar{w}) e^{\pm i m\hat p^\nu X_\nu},
\end{align}
from which the Minkowski vector indices are extracted in the same way as in (\ref{GEanyS}).

%%%%%%%%%%%%%%%%%%%%%%%%%%%%%%%%%%%%%%%%%%%%%%%%%

\subsection{Completeness of massive spinning conformal primary wave function basis}
Making use of the formalism explained in appendix \ref{op}, and employing eqs. (2.5), (3.10) and (3.29) from \cite{Costa:2011mg} in conjunction with eqs. (60), (83) and (93) from \cite{Costa:2014kfa} we have the identity
\begin{align}
\label{Gortho}
\sum_{J=-s}^s \int d\nu \mu_{s,J}(\nu)\int  \frac{d wd\bar{w}}{2^{|J|}}G^{(s)}_{J,1+i\nu}(\hat{p}_1; Y_1; w,\bar{w})  G^{(s)}_{-J,1-i\nu}(\hat{p}_2,Y_2;w,\bar{w})=\delta(\hat{p}_1,\hat{p}_2)(Y_{1}\cdot Y_{2})^s,
\end{align}
with integration weight (which includes normalizations dropped in (\ref{bulk to bdy}))
\begin{align}
 \mu_{s,J}(\nu) =(-1)^{|J|}\frac{J^2+ \nu^2}{4\pi^{3}}\frac{2^{s-|J|} (|J|+1)_{s-|J|}(\frac{1}{2}+|J|)_{s-|J|}}{(s-|J|)! (2|J|+1)_{s-|J|}} .
\end{align}
The delta function $\delta(\hat p_1,\hat p_2)$ has support when the two bulk points coincide
\begin{align}
\int_0^\infty \frac{dy_2}{y^3_2}\int dz_2 d\bar z_2 \delta(\hat p_1,\hat p_2)F(\hat p_1,\hat p_2)=F(\hat p_1,\hat p_1).
\end{align} 
With the help of (\ref{Gortho}) we can write down an inverse transform, taking a spin-$s$ massive conformal primary wave function back to the plane wave basis in Minkowski space
\begin{align}
&\epsilon_a^{\mu_1 ...\mu_s} \tilde Y_{\mu_1} ... \tilde Y_{\mu_s}  e^{\pm i m\hat p\,\cdot X}=\\
=&\frac{1}{s!(\frac{1}{2})_s}\epsilon_a^{\mu_1 ...\mu_s}K_{\mu_1}... K_{\mu_s}\sum_{J=-s}^s \int d\nu  \mu_{s,J}(\nu)\int \frac{d wd\bar{w}}{2^{|J|}}G^{(s)}_{-J,1-i\nu}(\hat{p}_1,Y; w,\bar{w})\phi^{\pm,m,s}_{J,1+i\nu}(X,\tilde{Y} ;w,\bar{w}).\notag
\end{align}
This demonstrates that the map to the celestial sphere is properly invertible, and the spinning massive conformal primary wave functions form a complete basis.

\section{Spin-$s$ massive momentum representation and Poincar\'e algebra}
\label{sec:smP}
As already seen in the massive scalar case \cite{Law:2019glh}, massive momentum operators on the celestial sphere are spin dependent. In fact, since a definite spin conformal primary wave function for a massive particle mixes all the spins in the bulk multiplet (\ref{phiAnyS}), the massive spin-$s$ momentum representation on the celestial sphere is simultaneously a Minkowski vector, as well as a matrix in spin multiplet space\footnote{The same is formally true in the massive scalar case, where $P_{J,I}^{\mu}$ with $\mu=0,1,2,3$ and $J,I=0$.}
\begin{align}
P_{J,I}^{\mu}~~~\text{with}~~~\mu=0,1,2,3\,,\text{ and }J,I=-s,-s+1,...,s-1,s\,.
\end{align}
The defining eigenvalue problem for the spin-$s$ momentum operator is
\begin{align}
\label{s1PphiEq}
\sum_{I=-s}^s P_{J,I}^{\mu}\sum_b G^{(s)}_{Ib}\epsilon_b^{\nu_1...\nu_s}=m\hat p^\mu\,\sum_b G^{(s)}_{Jb}\epsilon_b^{\nu_1...\nu_s},
\end{align}
involving the integration weight matrix $G^{(s)}_{Ja}$ of the spin-$s$ massive conformal primary wave function (\ref{phiAnyS}). Each component $\mu$ of the operator $P_{J,I}^{\mu}$ turns out to be a non-diagonal matrix in spin multiplet space, and we found it to be explicitly given by:
\begin{align}
\label{mPanyS}
&P_{J,I}^{\mu}=\frac{m}{2}\left[\frac{(J+s)  \delta_{J,I+1}}{ \Delta-J-1}\left((\partial_w q^\mu)+\frac{1}{\Delta+I} q^\mu\partial_{ w}\right)+\frac{(J-s)  \delta_{J+1,I}}{
   \Delta+J-1}\left((\partial_{\bar w} q^\mu)+\frac{1}{\Delta-I} q^\mu\partial_{\bar w}\right)\right.\notag\\
&\left.+\delta_{J,I}\left(\left((\partial_w\partial_{\bar w}q^\mu)+\frac{(\partial_{\bar w} q^\mu)\partial_{ w}}{\Delta-1+J}+\frac{(\partial_w q^\mu)\partial_{\bar w}}{\Delta-1-J}+ \frac{q^\mu\partial_{ w}\partial_{\bar w}}{(\Delta-1)^2-J^2}\right)e^{-\partial_\Delta }+\frac{(\Delta-s-1)(\Delta+s)}{{(\Delta-1)^2-J^2}} q^\mu e^{\partial_\Delta }\right)\right].
\end{align}
Note that since only $\delta_{J,I+1},\delta_{J,I},\delta_{J+1,I}$ are involved, only neighboring spins are coupled.

Since the normalization of each row in the integration weight matrix $G^{(s)}_{Ja}$  can be chosen separately, the off-diagonal terms of the massive momentum operator depend on this normalization as well. The result (\ref{mPanyS}) is designed to specifically match the row normalization for all $J$ explicitly chosen in (\ref{GEanyS}). 

It is straightforward to explicitly verify that this momentum operator indeed satisfies the defining equation (\ref{s1PphiEq}), e.g., for $s=0,1,2,3$ making use of (\ref{GEanyS}). Additionally, it is easy to analytically verify that it properly squares to $-m^2$
\begin{align}
\label{PPm2}
\sum_{K}\eta_{\mu\nu}P_{J,K}^{\mu}P_{K,I}^{\nu}=-m^2 \delta_{J,I},
\end{align}
for arbitrary integer spin $s\geq 0$.

All Lorentz generators $M_{J,I}^{\mu\nu}=-M_{J,I}^{\nu\mu}$ are diagonal in spin multiplet space (see \cite{Stieberger:2018onx})
\begin{align}
M_{J,I}^{01}&=\frac{i}{2} \delta_{J,I} \left(\left(\bar{w}^2-1\right)
   \partial_{\bar w}+\left(w^2-1\right)
   \partial_{w}+(\Delta-J) \bar{w}+(\Delta+J) w
   \right)\,,\\
	M_{J,I}^{02}&=\frac{-1}{2} \delta_{J,I} \left(\left(\bar{w}^2+1\right)
   \partial_{\bar w}-\left(w^2+1\right)
   \partial_{w}+(\Delta-J) \bar{w}-(\Delta+J) w
   \right)\,,\\
	M_{J,I}^{03}&=i\delta_{J,I} \left(\bar{w}  \partial_{\bar w}+w
   \partial_w+\Delta
   \right) ~~~~~~,~~~~~~
	M_{J,I}^{12}=\delta_{J,I}\left(-\bar{w}  \partial_{\bar w}+w
   \partial_w+J\right)
   \,,\\
	M_{J,I}^{13}&=\frac{i}{2} \delta_{J,I} \left(\left(\bar{w}^2+1\right)
   \partial_{\bar w}+\left(w^2+1\right)
   \partial_{w}+(\Delta-J) \bar{w}+(\Delta+J) w
   \right)\,,\\
	M_{J,I}^{23}&=\frac{-1}{2} \delta_{J,I} \left(\left(\bar{w}^2-1\right)
   \partial_{\bar w}-\left(w^2-1\right)
   \partial_{w}+(\Delta-J) \bar{w}-(\Delta+J) w
   \right)\,,
\end{align}
where $\partial_x=\frac{\partial}{\partial x}$. The Lorentz algebra can be written with spin multiplet indices made explicit
\begin{align}
[M^{\mu\nu},M^{\rho\sigma}]&=\sum_K\left(M_{J,K}^{\mu\nu}M_{K,I}^{\rho\sigma}-M_{J,K}^{\rho\sigma}M_{K,I}^{\mu\nu}\right)\notag\\
\label{LorentzAlgebra}
&=-i\left(\eta^{\mu\sigma}M_{J,I}^{\nu\rho}+\eta^{\nu\rho}M_{J,I}^{\mu\sigma}-\eta^{\mu\rho}M_{J,I}^{\nu\sigma}-\eta^{\nu\sigma}M_{J,I}^{\mu\rho}\right),
\end{align}
where $\eta^{\mu\nu}$ is the mostly plus Minkowski metric. The momentum operator $P_{J,I}^\mu$ then completes the Poincar\'e algebra
\begin{align}
\label{PoincareAlgebra}
\begin{split}
[M^{\mu\nu},P^{\rho}]&=\sum_K\left(M_{J,K}^{\mu\nu}P_{K,I}^{\rho}-P_{J,K}^{\rho}M_{K,I}^{\mu\nu}\right)=i\left(\eta^{\mu\rho}P_{J,I}^{\nu}-\eta^{\nu\rho}P_{J,I}^{\mu}\right)\,,\\
[P^{\mu},P^{\nu}]&=\sum_K\left(P_{J,K}^{\mu}P_{K,I}^{\nu}-P_{J,K}^{\nu}P_{K,I}^{\mu}\right)=0\,.
\end{split}
\end{align}
The proof that (\ref{mPanyS}) indeed is a correct (general spin $s$) massive momentum operator representation on the celestial sphere, is due to the fact that it squares to $-m^2$ as pointed out in (\ref{PPm2}) and properly closes the Poincar\'e algebra (\ref{PoincareAlgebra}) for arbitrary integer spin parameter values $s\geq 0$, which can be straightforwardly verified analytically.

The massless momentum representation for any spin $s$ is given by \cite{Stieberger:2018onx}
\begin{align}
\label{PanyS}
P_{J,I}^{\mu}=\delta_{J,I}q^\mu e^{\partial_\Delta},
\end{align}
which is diagonal in helicity space, same as the Lorentz generators $M_{J,I}^{\mu\nu}$, so that in the massless case the Poincar\'e algebra closes for each helicity independently and the spin multiplet indices can be suppressed.

\section{Poincar\'e constraints on massive spinning celestial amplitudes}
\label{sec:234constr}
As in the case of celestial amplitudes involving massive scalars \cite{Law:2019glh}, Lorentz symmetry constraints, with $i$ labeling the scattering particles from $1$ to $n$:
\begin{align}
\sum_{i=1}^n {M_i}^{\mu\nu}A=\sum_{i=1}^n \delta_{J_1,I_1}...\delta_{J_{i-1},I_{i-1}}{M_i}^{\mu\nu}_{J_i,I_i}\delta_{J_{i+1},I_{i+1}}...\delta_{J_n,I_n}A_{I_1,...,I_n}=0\,
\end{align}
with summation over each $-s_j\leq I_j\leq s_j$ for $j=1,2,...,n$ implied, are equivalent with conformal covariance of primary field correlator structures in CFT, so that the shape of each structure can be fixed to the following familiar form\footnote{Here and throughout the paper we sometimes use abbreviations $x_{ij}\equiv x_i-x_j$ for various quantities with particle indices.} \cite{Osborn:2012vt,DiFrancesco:1997nk}
\begin{align}
\label{A2ans}
A_2=&C^{J_{1},J_{2}}_{\Delta_1,\Delta_2}\frac{\delta(\Delta_{1}-\Delta_{2})\delta_{J_1,J_2}}{w_{12}^{h_{1}+h_{2}}\bar w_{12}^{\bar{h}_1+\bar{h}_2}},\\
\label{A3ans}
A_3=&C_{\Delta_1,\Delta_2,\Delta_3}^{J_{1},J_{2},J_{3}}w_{12}^{h_{3}-h_{1}-h_{2}}w_{23}^{h_{1}-h_{2}-h_{3}}w_{31}^{h_{2}-h_{1}-h_{3}}
 \bar w_{12}^{\bar{h}_3-\bar{h}_1-\bar{h}_2}\bar w_{23}^{\bar{h}_1-\bar{h}_2-\bar{h}_3}\bar w_{31}^{\bar{h}_2-\bar{h}_1-\bar{h}_3},\\
\label{A4ans}
A_4=&\frac{\left(\frac{w_{14}}{w_{13}}\right)^{h_{3}-h_{4}}
   \left(\frac{w_{24}}{w_{14}}\right)^{h_{1}-h_{2}}}{w_{12}^{h_{1}+h_{2}} w_{34}^{h_{3}+h_{4}}} 
   \frac{\left(\frac{\bar{w}_{14}}{\bar{w}_{13}}\right)^{\bar{h}_3-\bar{h}_4}
   \left(\frac{\bar{w}_{24}}{\bar{w}_{14}}\right)^{\bar{h}_1-\bar{h}_2}}{\bar{w}_{12}^{\bar{h}_1+\bar{h}_2}\bar{w}_{34}^{\bar{h}_3+\bar{h}_4}}f_{\Delta_1,\Delta_2,\Delta_3,\Delta_4}^{J_{1},J_{2},J_{3},J_{4}}(z ,\bar z),
\end{align} 
where in the two-point case we use Dirac and Kronecker deltas to keep $\Delta_1,\Delta_2$ and $J_1,J_2$ formally apart (which is required, since momentum operators act on particles 1 and 2 separately), and we use the conformal weight abbreviations
\begin{align}
h=\frac{\Delta+J}{2}~~~,~~~\bar h=\frac{\Delta-J}{2},
\end{align}
with conformal dimension $\Delta$ and spin $J$. In (\ref{A4ans}) we use the conformal cross-ratios
\begin{align}
\label{confcross}
z=\frac{w_{12}w_{34}}{ w_{13}w_{24}}~~~,~~~\bar z=\frac{\bar w_{12}\bar w_{34}}{ \bar w_{13}\bar w_{24}}.
\end{align}
Further constraints on the two- and three-point function coefficients $C$ and the function of cross-ratio $f$ are expected to result from momentum conservation requirements
\begin{align}
\sum_i \epsilon_i{P_i}^{\mu}A=\sum_{i=1}^n \epsilon_i\delta_{J_1,I_1}...\delta_{J_{i-1},I_{i-1}}{P_i}^{\mu}_{J_i,I_i}\delta_{J_{i+1},I_{i+1}}...\delta_{J_n,I_n}A_{I_1,...,I_n}=0,
\end{align}
where $\epsilon_i=\pm1$ for particle $i$ incoming/outgoing, and summation over each $-s_j\leq I_j\leq s_j$ for $j=1,2,...,n$ is implied.

In the following we systematically derive the resulting constraints.

\subsection{Two-point structure}
The two-point structure is special due to the presence of the delta functions $\delta(\Delta_{1}-\Delta_{2})\delta_{J_1,J_2}$. Application of the massive momentum operators in the equation $\sum_i \epsilon_i{P_i}^{\mu}A=0$ leads to four independent delta function combinations that cannot cancel each other.

On the support of $\delta(1+\Delta_{1}-\Delta_{2})\delta_{J_1,J_2}$ we have the constraint
\begin{align}
0=m_1 \epsilon _1 (s_1-\Delta +1) (\Delta +s_1)
   C^{{J,J}}_{{1 + \Delta ,1 + \Delta
   }}+m_2 \epsilon _2 (J-\Delta +1) (J+\Delta -1)
   C^{{J,J}}_{{\Delta ,\Delta }},
\end{align}
where we use the abbreviations $\Delta_2=\Delta+1,\,\Delta_1=\Delta$ and $J_1=J_2=J$. While, on the support of $\delta(-1+\Delta_{1}-\Delta_{2})\delta_{J_1,J_2}$ we obtain
\begin{align}
0=m_2
   \epsilon _2 \left(s_2-\Delta +1\right) \left(\Delta
   +s_2\right) C^{J,J}_{1 + \Delta ,1 +
   \Delta }+m_1 \epsilon _1 (J-\Delta+1) (J+\Delta -1)
   C^{{J,J}}_{\Delta ,\Delta },
\end{align}
where we use the abbreviations $\Delta_2=\Delta+1,\,\Delta_1=\Delta$ and $J_1=J_2=J$. To reconcile these two equations we must  take
\begin{align}
m_1=m_2=m~~~,~~~s_1=s_2=s\,,
\end{align}
while $\epsilon_1=-\epsilon_2$ is required by kinematics. Without loss of generality we rescale the two-point function coefficient as
\begin{align}
C_{\Delta,\Delta}^{J,J}=\frac{  \Gamma (\Delta-J -1) \Gamma
   (J+\Delta -1)}{\Gamma (\Delta-s -1) \Gamma (s+\Delta
   )}\left(-\frac{\epsilon _1}{\epsilon
   _2}\right)^{\Delta }c^J_\Delta,
\end{align}
such that the constraint above implies the periodicity of $c^J_\Delta$ under $\Delta\to\Delta+1$. Making use of this ansatz, the two remaining constraints on the support of $\delta(\Delta_{1}-\Delta_{2})\delta_{1+J_1,J_2}$ or $\delta(\Delta_{1}-\Delta_{2})\delta_{J_1,J_2+1}$ both reduce to
\begin{align}
0=\epsilon _1  (J+s+1) c^J_\Delta+\epsilon _2  (s-J) c^{J+1}_\Delta.
\end{align}
Considering that $J$ is integer, this equation has a known explicit recurrence solution. Formally reintroducing the distinction between $\Delta_1$ and $\Delta_2$ as well as $J_1$ and $J_2$, the general solution for the two-point function coefficient can be summarized as
\begin{align}
C_{\Delta_1,\Delta_2}^{J_1,J_2}=\frac{ \Gamma \left(h_1+h_2-1\right)
   \Gamma \left(\bar{h}_1+\bar{h}_2-1\right) \Gamma
   \left(s-\frac{J_1}{2}-\frac{J_2}{2}+1\right) \Gamma
   \left(s+\frac{J_1}{2}+\frac{J_2}{2}+1\right)}{\Gamma
   \left(-s+\frac{\Delta _1}{2}+\frac{\Delta
   _2}{2}-1\right) \Gamma \left(s+\frac{\Delta
   _1}{2}+\frac{\Delta _2}{2}\right)}\left(-\frac{\epsilon_1 }{\epsilon_2}\right)^{h_1+h_2}c_{\frac{\Delta_1+\Delta_2}{2}},
\end{align}
where $c_x$ is periodic under $x\to x+1$, and we kept the spin multiplet number $s_1=s_2=s$ the same. In case $s=J_1=J_2=0$ this readily reduces to the scalar result found in \cite{Law:2019glh}. Note that, as discussed in \cite{Law:2019glh}, in order to avoid convergence issues in the inverse integral transform back to Minkowski space, $c_x$ better be trivially periodic (ergo constant). However, again as in the scalar case \cite{Law:2019glh}, even without periodicities the inverse transform to Minkowski space is not defined, since the Mellin-like integration over $\Delta_1=\Delta_2=\Delta=1+i\nu$ with $-\infty< \nu<\infty$ directly hits a pole at $\nu=0$ in the (pseudo-) scalar contribution of the multiplet $J_1=J_2=J=0$. Therefore, even though a non-trivial two-point celestial structure exists, its map to Minkowski space is not well defined. This is consistent with the fact that there is no non-trivial two-point amplitude in Minkowksi space.

The two-point function coefficient for a two-point structure of one massless and one massive particle is constrained to vanish, same as in the scalar case.

\subsection{Three-point structure}
In the three point case we can consider amplitudes involving one, two, or three massive external particles, while the remaining particles are massless. The case when the massive particles are scalars has been investigated previously \cite{Law:2019glh}. In this section we repeat the analysis for massive particles of non-zero spin.

\paragraph{Two massless, one massive:} When one of the three particles is a spinning massive boson, the equations $\sum_i \epsilon_i{P_i}^{\mu}A^{m,0,0}=0$ lead to four independent constraints, which can be expressed symmetrically in indices $2$ and $3$ for overall upper and lower cases of the double-signs as:\footnote{To obtain these constraints, calculate the action of the momenta operators, divide out the common $w_i,\bar w_i$ dependence and require the vanishing of all coefficients of the remaining different $w_i,\bar w_i$ monomials.}
\begin{align}
\label{A00mR1}
&0=  \frac{m}{4} \epsilon _1 \left(\left(\Delta _1\pm\Delta _2\mp\Delta
   _3\pm J_1+J_2-J_3-1\right) \left(\Delta _1\pm\Delta
   _2\mp\Delta _3\mp J_1-J_2+J_3-1\right)
   C_{\Delta _1-1,\Delta
   _2,\Delta
   _3}^{J_1,J_2,J_3}\right.\notag\\%
	&\left.-
   \left(\Delta _1-s_1-1\right) \left(\Delta
   _1+s_1\right)
   C_{\Delta _1+1,\Delta
   _2,\Delta
   _3}^{J_1,J_2,J_3}\right)\mp 2
   \epsilon _2 \left(J_1
   \left(J_2-J_3\right)-\left(\Delta _1-1\right)
   \left(\Delta _2-\Delta _3\right)\right)
   C_{\Delta _1,\Delta
   _{2}+\frac{1\pm1}{2},\Delta
   _{3}+\frac{1\mp1}{2}}^{J_1,J_2,J_3},\\
\label{A00mR3}
0&=4 \left(J_1 \left(J_2-J_3\right)-\left(\Delta _1-1\right)
   \left(\Delta _2-\Delta _3\right)\right)
   \left(J_1\mp s_1\right)
   C_{\Delta _1,\Delta
   _2,\Delta
   _3}^{J_1\pm 1,J_2,J_3}\notag\\
	&-4
   \left(\Delta _1\pm J_1-1\right) \left(\Delta
   _1-s_1-1\right) \left(\Delta _1+s_1\right)
   C_{\Delta _1+1,\Delta
   _2,\Delta
   _3}^{J_1,J_2,J_3}\\
	&+\left(\Delta _1\mp J_1-1\right) \left(\Delta _1\pm \Delta _2 \mp\Delta
   _3\pm J_1+J_2-J_3-1\right) \left(\Delta _1\mp\Delta
   _2\pm\Delta _3\pm J_1\mp J_2+J_3-1\right)
   C_{\Delta _1-1,\Delta
   _2,\Delta _3}^{J_1,J_2,J_3}.\notag
\end{align}
All four of these recurrence relations involve three different terms with non-diagonal parameter shifts which makes this system of equations hard to solve in general. However, note that one of the two equations (\ref{A00mR3}) reduces to two-term recurrence relations with shifts in the parameter $\Delta_1$ only, in the case when $J_1=s_1$ or $J_1=-s_1$ respectively. With this it is possible to make progress as follows.

If we start, e.g., with the highest weight $J_1=s_1$, then without loss of generality we choose to rescale the respective three-point function coefficient as
\begin{align}
\label{C3m00ans}
C_{\Delta_1,\Delta_2,\Delta_3}^{s_1,J_2,J_3}=\frac{ \left(-\frac{m \epsilon _1}{2\epsilon
   _2}\right)^{\Delta _2} \left(-\frac{m \epsilon _1}{2\epsilon
   _3}\right)^{\Delta _3}  \Gamma \left(\frac{s_1+J_2-J_3+\Delta
   _1+\Delta _2-\Delta _3}{2} \right) \Gamma \left(\frac{s_1-J_2+J_3+\Delta _1-\Delta _2+\Delta _3}{2}
   \right)}{\Gamma
   \left(s_1+\Delta _1\right)}c_{\Delta_1,\Delta_2,\Delta_3}^{s_1,J_2,J_3},
\end{align}
where, kinematically, only the incoming-outgoing combination $\epsilon_1=-\epsilon_2=-\epsilon_3$ makes sense. With this, the upper sign equation in (\ref{A00mR3}) reduces to
\begin{align}
c_{\Delta_1-1,\Delta_2,\Delta_3}^{s_1,J_2,J_3}=c_{\Delta_1+1,\Delta_2,\Delta_3}^{s_1,J_2,J_3},
\end{align}
implying that $c$ must be periodic under $\Delta_1\to\Delta_1+2$.
 Using this, the equations (\ref{A00mR1}) with $J_1=s_1$ simplify to
\begin{align}
c_{\Delta_1+1,\Delta_2,\Delta_3}^{s_1,J_2,J_3}=c_{\Delta_1,\Delta_2+1,\Delta_3}^{s_1,J_2,J_3}~~~\text{and}~~~c_{\Delta_1+1,\Delta_2,\Delta_3}^{s_1,J_2,J_3}=c_{\Delta_1,\Delta_2,\Delta_3+1}^{s_1,J_2,J_3}.
\end{align}
As explained in appendix B of \cite{Law:2019glh}, this implies that $\Delta_i$ dependence in $c$ must either appear in combination $\sum_i\Delta_i$ (in which case it must be periodic under $\sum_i\Delta_i\to \sum_i\Delta_i+2$ due to periodicity of $\Delta_1$), or else must be periodic under $\Delta_i\to\Delta_i+1$. For the same reason as discussed at the end of section 4 in \cite{Law:2019glh}, assuming on physical grounds that $c$ is a holomorphic function of finite order, a non-trivial periodicity of $c$ makes it unbounded on any complex vertical line in the complex plane. Therefore, to ensure convergence of the inverse integral transform of the celestial amplitude back to Minkowski space, $c$ better be trivially periodic -- ergo constant.\footnote{Considering $\Delta_i=1+i\lambda_i$, the ratio of gamma functions in (\ref{C3m00ans}) vanishes as $\sqrt{1/\lambda_1}$ for $\lambda_1\to\pm \infty$ and as $e^{-\pi|\lambda_j|/2}$ for $\lambda_j\to\pm \infty$ with $j=2,3$. Therefore, even the first non-trivial periodic mode $e^{2\pi i n \Delta}$ in a Laureant expansion of an integer-periodic holomorphic function would dominate the suppressing effect of the gamma function ratio and must therefore be ruled out.}

The ansatz (\ref{C3m00ans}) as well as the periodicities of $c$ can then be used in the lower sign equation in (\ref{A00mR3}) to iteratively generate $C_{\Delta_1,\Delta_2,\Delta_3}^{s_1-l,J_2,J_3}$ for $l=1,2,...,2s_1$ in terms of the initial $J_1=s_1$ expression. The iteration must truncate after $l=2s_1$ steps, at which point the $J_1$ spin multiplet is complete. A factor of $(J_1+s_1)$ in (\ref{A00mR3}) ensures that no further $C_{\Delta _1,\Delta
   _2,\Delta
   _3}^{J_1-1,J_2,J_3}$ term is generated once $J_1=-s_1$; however, the rest of the terms in this equation must then cancel exactly for the multiplet truncation to be consistent. Explicitly generating multiplets for a few $s_1$ values such as $s_1=1$, $s_1=2$, or $s_1=3$, reveals that the rest of the terms in the $J_1=-s_1$ truncation equation does not cancel automatically, but instead becomes proportional to
\begin{align}
\propto\prod_{j=-s_1}^{s_1}(J_2-J_3+j).
\end{align}
The consistency requirement of the multiplet truncation therefore implies that the difference of helicity values of two massless particles produced in the decay of a massive particle may never exceed the maximum spin that can be accommodated by the spin multiplet of the massive particle\footnote{Compare with section 4.1 in \cite{Arkani-Hamed:2017jhn}.}
\begin{align}
|J_2-J_3|\leq s_1.
\end{align}
Naturally, we can start with $J_1=-s_1$ and build up the multiplet in the opposite direction instead. In that case the role of eqs. (\ref{A00mR3}) with upper and lower signs is reversed, but all multiplet construction steps proceed analogously.

In appendix \ref{sec:3pexamples} we consider explicit examples of three-point amplitudes with two massless scalars and one massive spinning external particle mapped to the celestial sphere. There we find exact agreement with the three-point function coefficient obtained above purely from symmetry, with $c$ trivially periodic in conformal dimensions $c_{\Delta_1,\Delta_2,\Delta_3}^{s_1,J_2,J_3}=const$. 

\paragraph{One massless, two massive:} When two of the three particles are spinning massive bosons, the equations $\sum_i \epsilon_i{P_i}^{\mu}A^{m_1,m_2,0}=0$ lead to four independent constraints, which can be expressed symmetrically in indices $1$ and $2$ as:
\begin{align}
\label{Amm0R1}
0=&4 \epsilon _3
   C_{\Delta _1,\Delta
   _2,\Delta
   _3+1}^{J_1,J_2,J_3}\\
&+\frac{m_2 \epsilon _2 \left(2
   \left(J_2\mp s_2\right)
   C_{\Delta _1,\Delta
   _2,\Delta
   _3}^{J_1,J_2\pm 1,J_3}+\left(-
   \Delta _1+\Delta _2+\Delta _3\mp J_1\pm J_2\pm J_3-1\right)
   C_{\Delta _1,\Delta
   _2-1,\Delta
   _3}^{J_1,J_2,J_3}\right)}{\Delta _2\pm J_2-1}\notag\\
&+\frac{m_1 \epsilon _1 \left(2 \left(J_1\pm s_1\right)
   C_{\Delta _1,\Delta
   _2,\Delta
   _3}^{J_1\mp 1,J_2,J_3}+\left(-
   \Delta _1+\Delta _2-\Delta _3\pm J_1\mp J_2\pm J_3+1\right)
   C_{\Delta _1-1,\Delta
   _2,\Delta
   _3}^{J_1,J_2,J_3}\right)}{-
   \Delta _1\pm J_1+1}\,,\notag\\
\label{Amm0R3}
0=&\frac{m_i \epsilon _i
   \left(-\Delta _i+s_i+1\right) \left(\Delta
   _i+s_i\right)
   e^{\partial_{\Delta_i}}C_{\Delta _1,\Delta
   _2,\Delta
   _3}^{J_1,J_2,J_3}}{-\Delta
   _i+J_i+1}+\epsilon _3 \left(\Delta
   _1+\Delta _2-\Delta _3+J_1+J_2-J_3-1\right)
   C_{\Delta _1,\Delta
   _2,\Delta
   _3+1}^{J_1,J_2,J_3}\notag\\
&+\frac{m_j \epsilon _j \left(\Delta _1+\Delta
   _2-\Delta _3+J_1+J_2-J_3-1\right) \left(\Delta
   _1+\Delta _2+\Delta _3+J_1+J_2+J_3-3\right)
   e^{-\partial_{\Delta_j}}C_{\Delta _1,\Delta
   _2,\Delta
   _3}^{J_1,J_2,J_3}}{4
   \left(\Delta _j+J_j-1\right)}\notag\\
&+\frac{m_2 \epsilon _2 \left(\mp\Delta _1\pm\Delta _2-\Delta
   _3-J_1+J_2\mp J_3+1\right) \left(J_2-s_2\right)
   C_{\Delta _1,\Delta
   _2,\Delta
   _3}^{J_1,J_2+1,J_3}}{2
   \left(\pm\Delta _2+J_2\mp 1\right)}\\
&-\frac{m_1 \epsilon _1
   \left(\mp\Delta _1\pm\Delta _2-\Delta
   _3+J_1-J_2\pm J_3+1\right) \left(J_1-s_1\right)
   C_{\Delta _1,\Delta
   _2,\Delta
   _3}^{J_1+1,J_2,J_3}}{2
   \left(\mp\Delta _1+J_1\pm 1\right)},\notag
\end{align}
where in (\ref{Amm0R3}) the index combination $i=1,j=2$ is taken together with the upper of the double-signs, and combination $i=2,j=1$ is taken with the lower of the double-signs. 

\paragraph{Zero massless, three massive:} When all three particles are spinning massive bosons, the equations $\sum_i \epsilon_i{P_i}^{\mu}A^{m_1,m_2,m_3}=0$ again lead to four independent constraints. One constraint can be given as:
\begin{align}
0=&\left(\left(\frac{ m_1 \epsilon _1 \left(s_1-J_1\right) }{\Delta
   _1+J_1-1}C_{\Delta
   _1,\Delta _2,\Delta _3}^{J_1+1,J_2,J_3}+\frac{ m_1 \epsilon _1 \left(J_1+s_1\right)
   }{\Delta _1-J_1-1}C_{\Delta _1,\Delta _2,\Delta
   _3}^{J_1-1,J_2,J_3}\right)+(cyclic)\right)\\
	&+\left(\left(\frac{ m_1 \epsilon _1 \left(\left(\Delta _1-1\right)
   J_2+\left(\Delta _3-\Delta _2\right) J_1+(1-\Delta _1) J_3\right)
   }{J_1^2-\left(\Delta _1-1\right){}^2}C_{\Delta _1-1,\Delta _2,\Delta
   _3}^{J_1,J_2,J_3}\right)+(cyclic)\right),\notag
\end{align}
where $(cyclic)$ denotes terms with indices $1,2,3$ cyclically interchanged. The remaining three constraints can be summarized as (abbreviating $C=C^{J_1,J_2,J_3}_{\Delta _1,\Delta _2,\Delta
   _3}$):
\begin{align}
0=&\frac{m_i \epsilon _i \left(1-2 h_i-2 h_j+2 h_k\right) \left(1-2
   \bar{h}_i+2 \bar{h}_j-2 \bar{h}_k\right)
   }{\left(1-2 h_i\right) \left(1-2 \bar{h}_i\right)}e^{-\partial_{\Delta_i}}C-\frac{2
   m_k \epsilon _k \left(1-2 \bar{h}_i+2 \bar{h}_j-2 \bar{h}_k\right)
   }{1-2 \bar{h}_k}e^{-\partial_{\Delta_k}}C\notag\\
\begin{split}
&+\frac{4 m_i \epsilon _i \left(1-\Delta
   _i+s_i\right) \left(\Delta _i+s_i\right)
   }{\left(1-2 h_i\right) \left(1-2 \bar{h}_i\right)}e^{\partial_{\Delta_i}}C-\frac{2
   m_j \epsilon _j \left(1-2 h_i-2 h_j+2 h_k\right)
   }{1-2 h_j}e^{-\partial_{\Delta_j}}C\\
&+\left(\frac{4
   m_k \epsilon _k \left(J_k+s_k\right)
   }{1-2 \bar{h}_k}S_{J_k}^{-1}C+\frac{4 m_j \epsilon _j
   \left(J_j-s_j\right) }{2 h_j-1}S_{J_j}^{+1}C\right.
\end{split}\\
&\left.+\frac{2 m_i \epsilon _i \left(J_i-s_i\right) \left(1-2 \bar{h}_i+2
   \bar{h}_j-2 \bar{h}_k\right) }{\left(2 h_i-1\right) \left(1-2
   \bar{h}_i\right)}S_{J_i}^{+1}C+\frac{2 m_i \epsilon _i \left(J_i+s_i\right)
   \left(2 h_i+2 h_j-2 h_k-1\right)
   }{\left(2 h_i-1\right) \left(1-2 \bar{h}_i\right)}S_{J_i}^{-1}C\right)\,,\notag
\end{align}
one equation for each index combination $(i,j,k)\in\{(1,2,3),(2,3,1),(3,1,2)\}$. Here we have defined a spin-shift operator $S_{J_i}^{\pm 1}$, which acts to the right as $S_{J_i}^{\pm 1}C^{...,J_i,...}_{...}=C^{...,J_i\pm 1,...}_{...}$.

\subsection{Four-point structure}

\paragraph{Three massless, one massive:}
Momentum conservation $\sum_i \epsilon_i{P_i}^{\mu}A^{0,0,0,m}=0$ in case of three massless particles and one mass $m$ particle of arbitrary spin leads to four constraint equations. The first constraint features only shifts in operator dimensions
\begin{align}
0=&\sqrt{z \bar{z}} \left(m_4 \epsilon _4
   f^{J_1,J_2,J_3,J_4}_{\Delta _1,\Delta _2,\Delta _3,\Delta
   _4-1}(z,\bar{z})+2 \epsilon _3 f^{J_1,J_2,J_3,J_4}_{\Delta
   _1,\Delta _2,\Delta _3+1,\Delta _4}(z,\bar{z})\right)\\
	&+2
   \epsilon _1 f^{J_1,J_2,J_3,J_4}_{\Delta _1+1,\Delta _2,\Delta
   _3,\Delta _4}(z,\bar{z})+2 \epsilon _2
   f^{J_1,J_2,J_3,J_4}_{\Delta _1,\Delta _2+1,\Delta _3,\Delta
   _4}(z,\bar{z}),\notag
\end{align}
where the conformal cross ratios $z,\bar z$ were defined in (\ref{confcross}). Two further constraints involve shifts in conformal dimensions, spin of the massive particle, and first order derivatives with respect to conformal cross ratios
\begin{align}
\label{constr4pt2}
0=&\frac{m \epsilon _4 \left(\Delta _3+\Delta _4+z \left(-\Delta _1+\Delta _2-J_1+J_2\right)+J_3+J_4-1\right) }{4 \left(\Delta _4+J_4-1\right)}f^{J_1,J_2,J_3,J_4}_{\Delta _1,\Delta _2,\Delta _3,\Delta _4-1}(z,\bar{z})\notag\\
&+\epsilon _3 f^{J_1,J_2,J_3,J_4}_{\Delta _1,\Delta _2,\Delta _3+1,\Delta _4}(z,\bar{z}) +
\sqrt{\frac{z  }{\bar{z}}}\epsilon _2f^{J_1,J_2,J_3,J_4}_{\Delta _1,\Delta_2+1,\Delta _3,\Delta _4}(z,\bar{z})\\
&-\frac{m \epsilon _4 \left(s_4-J_4\right) }{2 \left(\Delta _4+J_4-1\right)}f^{J_1,J_2,J_3,J_4+1}_{\Delta _1,\Delta _2,\Delta _3,\Delta _4}(z,\bar{z})+\frac{m (z-1) z \epsilon _4  }{2 \left(\Delta _4+J_4-1\right)}\partial_z f^{J_1,J_2,J_3,J_4}_{\Delta _1,\Delta _2,\Delta _3,\Delta _4-1}(z,\bar{z}),\notag
\end{align}
\begin{align}
\label{constr4pt3}
0=&\frac{m \epsilon _4 \left(\bar{z} \left(\Delta _1-\Delta _2-J_1+J_2\right)-\Delta _3-\Delta _4+J_3+J_4+1\right) }{4 \left(-\Delta _4+J_4+1\right)}f^{J_1,J_2,J_3,J_4}_{\Delta _1,\Delta _2,\Delta _3,\Delta _4-1}(z,\bar{z})\notag\\
&+\epsilon _3 f^{J_1,J_2,J_3,J_4}_{\Delta _1,\Delta _2,\Delta _3+1,\Delta _4}(z,\bar{z})+\sqrt{\frac{ \bar{z} }{z}}\epsilon _2f^{J_1,J_2,J_3,J_4}_{\Delta _1,\Delta _2+1,\Delta _3,\Delta _4}(z,\bar{z})\\
&+\frac{m \epsilon _4 \left(J_4+s_4\right) }{2 \left(\Delta _4-J_4-1\right)}f^{J_1,J_2,J_3,J_4-1}_{\Delta _1,\Delta _2,\Delta _3,\Delta _4}(z,\bar{z})+\frac{m \epsilon _4 \left(\bar{z}-1\right) \bar{z} }{2 \left(\Delta _4-J_4-1\right)}\partial_{\bar z}f^{J_1,J_2,J_3,J_4}_{\Delta _1,\Delta _2,\Delta _3,\Delta _4-1}(z,\bar{z}).\notag
\end{align}
The fourth constraint involves shifts in conformal dimensions, spin of the massive particle, as well as first and second order derivatives with respect to conformal cross ratios
\begin{align}
\label{constr4pt4}
0&=\frac{\epsilon _2  }{ \sqrt{z\bar{z}}}\left(\frac{ z \left(\bar{z} \left(\bar h _2- \bar h _1\right)+\bar h _3+\bar h _4-\frac{1}{2}\right)}{1-2\bar h _4}+\frac{\bar{z} \left(z \left(h _2-h _1\right)+h _3+h _4-\frac{1}{2}\right)}{1-2h _4}+ z \bar{z}\right)f^{J_1,J_2,J_3,J_4}_{\Delta _1,\Delta _2+1,\Delta _3,\Delta _4}(z,\bar{z})\notag\\
&+\epsilon _3 \left(\frac{ \left(\bar{h}_1-\bar{h}_2\right) \bar{z}- \bar{h}_3}{2 \bar{h}_4-1}+\frac{ \left(h_1  - h_2\right) z- h_3}{2 h_4-1}\right) f^{J_1,J_2,J_3,J_4}_{\Delta _1,\Delta _2,\Delta _3+1,\Delta _4}(z,\bar{z})\\
&+\frac{m \epsilon _4 \left(\Delta_4-s_4-1\right) \left(\Delta_4+s_4\right) }{2 \left(2 h_4-1\right) \left(2 \bar{h}_4-1\right)}f^{J_1,J_2,J_3,J_4}_{\Delta _1,\Delta _2,\Delta _3,\Delta _4+1}(z,\bar{z})+\frac{m \epsilon _4 (z-1) z  \left(\bar{z}-1\right) \bar{z} }{2 \left(2 h_4-1\right) \left(2 \bar{h}_4-1\right)}\partial_z \partial_{\bar z} f^{J_1,J_2,J_3,J_4}_{\Delta _1,\Delta _2,\Delta _3,\Delta _4-1}(z,\bar{z})\notag \\
&-\frac{m \epsilon _4 \left( (h_2 - h_1) z+ h_3+ h_4-\frac{1}{2}\right) \left( \left(\bar{h}_2-\bar{h}_1\right) \bar{z}+ \bar{h}_3+ \bar{h}_4-\frac{1}{2}\right) }{2 \left(2 h_4-1\right) \left(2 \bar{h}_4-1\right)}f^{J_1,J_2,J_3,J_4}_{\Delta _1,\Delta _2,\Delta _3,\Delta _4-1}(z,\bar{z})
\notag\\
&+\frac{m \epsilon _4 \left(\bar{z}-1\right) \bar{z} \left(J_4-s_4\right) }{2 \left(2 h_4-1\right) \left(2 \bar{h}_4-1\right)}\partial_{\bar z}f^{J_1,J_2,J_3,J_4+1}_{\Delta _1,\Delta _2,\Delta _3,\Delta _4}(z,\bar{z})
+\frac{m (z-1) z \epsilon _4 \left(J_4+s_4\right) }{2 \left(2 h_4-1\right) \left(2 \bar{h}_4-1\right)}\partial_z f^{J_1,J_2,J_3,J_4-1}_{\Delta _1,\Delta _2,\Delta _3,\Delta _4}(z,\bar{z}).\notag
\end{align}
In case when the massive particle is a scalar $s_4=J_4=0$, these equations properly reduce to the constraints found in \cite{Law:2019glh}.

\paragraph{Two massless, two massive and beyond:} As in the massive scalar case \cite{Law:2019glh}, for the four-point structure, momentum conservation always leads to four constraints which consist of coupled shift equations in conformal dimensions of the four-particles, while simultaneously being differential equations in conformal cross ratios. In the massive spinning case, shifts in spins of massive particles also occur. When two or more massive particles participate, the resulting four equations similarly can be straightforwardly generated from the application of momentum operators to (\ref{A4ans}), but are quite unwieldy. Therefore, we do not spell these equations out explicitly.

\acknowledgments
We thank F. Denef, C. Sleight and M. Taronna for interesting discussions. We are especially grateful to A. Joyce for in-depth conversations and useful suggestions throughout the course of this project. AL and MZ are supported by the US Department of Energy under
contract DE-SC0011941.

\appendix

\section{Spin-one and -two massive polarization tensors}
\label{s1s2polarizations}
Spin of massive particles is encoded by polarization tensors (compare this section, e.g., with section 2.2 in \cite{Hinterbichler:2017qyt}). In the spin-1 case the polarization vectors of spin $a=-1,0,+1$ describing circular polarizations in the plane perpendicular to momentum $m \hat p^{\mu}$ are parametrized as
\begin{align}
\label{s1pols}
\epsilon_{-1}^\mu=\sqrt{2}y\partial_{\bar{z}}\hat{p}^\mu~~~,~~~\epsilon_{0}^\mu=y\partial_{y}\hat{p}^\mu~~~,~~~\epsilon_{+1}^\mu=\sqrt{2}y\partial_{z}\hat{p}^\mu\,,
\end{align}
which satisfy all required polarization properties:
\begin{align}
\epsilon_{a}\cdot\hat{p}=0~~~,~~~\epsilon_{a}^*\cdot{\epsilon_{b}}=\delta_{a,b}~~~,~~~\sum_{a=-1}^1{\epsilon_{a}^*}^\mu{\epsilon_{a}}^\nu=\eta^{\mu\nu}-\frac{\hat p^\mu\hat p^\nu}{\hat p^2}.
\end{align}
Higher spin polarization tensors generically can be composed out of the spin-$1$ polarization vectors given above by taking appropriate symmetric and traceless combinations. For instance, in the spin-$2$ case the $a=-2,-1,0,+1,+2$ components read
\begin{align}
\label{s2pols}
{\epsilon_{-2}^{\mu\nu}=\epsilon_{-1}^\mu\epsilon_{-1}^\nu\,,~~~\epsilon_{-1}^{\mu\nu}=\frac{i}{\sqrt{2}}\left(\epsilon_{-1}^\mu\epsilon_{0}^\nu+\epsilon_{0}^\mu\epsilon_{-1}^\nu\right)\,,\above 0pt \epsilon_{+2}^{\mu\nu}=\epsilon_{+1}^\mu\epsilon_{+1}^\nu\,,~~~\epsilon_{+1}^{\mu\nu}=\frac{-i}{\sqrt{2}}\left(\epsilon_{+1}^\mu\epsilon_{0}^\nu+\epsilon_{0}^\mu\epsilon_{+1}^\nu\right)\,,}~~~\epsilon_{0}^{\mu\nu}=\sqrt{\frac{2}{3}}\left(\epsilon_{0}^\mu\epsilon_{0}^\nu-\frac{1}{2}\epsilon_{-1}^\mu\epsilon_{+1}^\nu-\frac{1}{2}\epsilon_{+1}^\mu\epsilon_{-1}^\nu\right),
\end{align}
such that once again all required polarization properties are satisfied:
\begin{align}
\epsilon_{a}^{\mu\nu}\hat{p}_\nu=0~,~~~\epsilon_{a}^*{}^{\mu\nu}{\epsilon_{b}}_{\mu\nu}=\delta_{a,b}~,~~~\sum_{a=-2}^2{\epsilon_{a}^*}^{\mu\nu}{\epsilon_{a}}^{\rho\sigma}=\frac{1}{2}P^{\mu\rho}P^{\nu\sigma}+\frac{1}{2}P^{\nu\rho}P^{\mu\sigma}-\frac{1}{3}P^{\mu\nu}P^{\rho\sigma},
\end{align}
where $P^{\mu\nu}=\eta^{\mu\nu}-\frac{\hat p^\mu\hat p^\nu}{\hat p^2}$, and $\epsilon_a^{\mu\nu}\eta_{\mu\nu}=0$ for each index $a$ by construction.

\section{Polynomial encoding of symmetric traceless transverse tensors review}\label{op}
In this appendix we review the formalism of encoding symmetric traceless transverse tensors in terms of polynomials of auxiliary variables \cite{Costa:2011mg}.
\subsection{Tensors living on $\hat{p}^2 +1=0$ in $\mathbb{R}^{1,3}$}

When we work with a spin-$s$ symmetric traceless and transverse (STT) tensor $H_{\mu_1 \cdots \mu_s }(\hat{p})$ living on the submanifold $\hat{p}^2 +1=0$ in $\mathbb{R}^{1,3}$, it is useful to express it in terms of a homogeneous polynomial in auxiliary variables $Y^\mu$
\begin{align}\label{hyperbolic poly}
H(\hat{p},Y)=H_{\mu_1 \cdots \mu_s }(\hat{p})Y^{\mu_1}\cdots Y^{\mu_s}.
\end{align}
As explained in \cite{Costa:2011mg}, the information of an STT tensor is fully encoded by this polynomial if the following restrictions are imposed on the auxiliary variable $Y$:
\begin{align}\label{hyperbolic submfd}
\hat{p}^2 +1=Y^2=\hat{p}\cdot Y =0.
\end{align}
We think of the polynomial \eqref{hyperbolic poly} as living on this sub-manifold. To extract the original STT tensor from $H(\hat{p},Y)$, we use the differential operator
\begin{align}\label{hyper recover op}
K_\mu =&\frac{1}{2}\left(\frac{\partial}{\partial Y^\mu}+\hat{p}_\mu\left(\hat{p}\cdot \frac{\partial}{\partial Y}\right) \right)+\left(Y\cdot \frac{\partial}{\partial Y} \right)\frac{\partial}{\partial Y^\mu}\\
&+\hat{p}_\mu\left(Y\cdot \frac{\partial}{\partial Y} \right)\left(\hat{p}\cdot \frac{\partial}{\partial Y} \right)-\frac{Y_\mu}{2}\left(\frac{\partial^2}{\partial Y\cdot \partial Y} +\left(\hat{p}\cdot \frac{\partial}{\partial Y} \right)\left(\hat{p}\cdot \frac{\partial}{\partial Y} \right)\right),\nonumber
\end{align}
which acts interior to \eqref{hyperbolic submfd}, i.e. for any arbitrary function $g(Y)$ we have 
\begin{align}
K_\mu((\hat{p}^2+1) g(Y))=K_\mu(Y^2 g(Y))=K_\mu(\hat{p}\cdot Y g(Y))=0.
\end{align}
Additionally, it is easy to verify that the operator is symmetric, transverse and traceless
\begin{align}
K_\mu K_\nu=K_\nu K_\mu,\qquad \hat{p}^\mu K_\mu=0, \qquad K_\mu K^\mu=0,
\end{align}
and therefore its action on any polynomial of $Y$ defines a transverse traceless symmetric tensor on  \eqref{hyperbolic submfd}. 
The original tensor can be recovered from the polynomial via
\begin{align}
H_{\mu_1 \cdots \mu_s }(\hat{p})=\frac{1}{s!\Big(\frac{1}{2} \Big)_s}K_{\mu_1}\cdots K_{\mu_s}H(\hat{p},Y),
\end{align}
where $(a)_n = \frac{\Gamma(a+n)}{\Gamma(a)}$ is the Pochhammer symbol. 

To take formal derivatives of an STT tensor with respect to $\hat{p}^\mu$, we use the covariant derivative operator
\begin{align}
 \nabla_{\hat{p}^\mu} =\frac{\partial}{\partial \hat{p}^\mu}+\hat{p}_\mu \left(\hat{p}\cdot\frac{\partial}{\partial \hat{p}} \right)+Y_\mu \left(\hat{p}\cdot\frac{\partial}{\partial Y} \right),
\end{align}
which acts interior to the submanifold  \eqref{hyperbolic submfd}, and transverse ($\hat{p}\cdot \nabla_{\hat{p}}=0$). This satisfies $Y\cdot \nabla_{\hat{p}}=Y\cdot \partial_{\hat{p}}$.

\subsection{Tensors living on $q^2=0$ in $\mathbb{R}^{1,3}$ and projection onto the boundary}\label{lc op}

On the two dimensional celestial sphere, any spin-$J$ symmetric traceless tensor $f_{a_1 \cdots a_J}(w,\bar{w})$ has only two independent components. In terms of complex coordinates $w,\bar{w}$, we consider the components
\begin{align}\label{d=2 ten}
f_J(w,\bar{w})= f_{w  \cdots w}(w,\bar{w})\quad \text{and} \quad \tilde{f}_J(w,\bar{w})=f_{\bar{w}\cdots \bar{w}}(w,\bar{w}),
\end{align}
since all other components are forced to vanish due to the tracelessness condition. We can think of \eqref{d=2 ten} as all plus or all minus spin components respectively, since they involve $w$ or $\bar{w}$ indices exclusively. However, these components are not complex conjugates of each other in general. 

In our case of interest, we are dealing with $f_J(w,\bar{w})$ and $\tilde{f}_J(w,\bar{w})$ defined as pull-backs to the celestial sphere from a $\mathbb{R}^{1,3}$ symmetric traceless tensor $ F _{\mu_1 \cdots \mu_J }(q)$ living on $q^2=0$, such that
\begin{align}\label{d=2 projection}
f_J(w,\bar{w})=\frac{\partial q^{\mu_1}}{\partial w}\cdots \frac{\partial q^{\mu_J}}{\partial w}  F_{\mu_1 \cdots \mu_J }(q)\quad\text{and}\quad \tilde{f}_J(w,\bar{w})=\frac{\partial q^{\mu_1}}{\partial \bar{w}}\cdots \frac{\partial q^{\mu_J}}{\partial \bar{w}}  F_{\mu_1 \cdots \mu_J }(q),
\end{align}
where $ F$ is transverse $q^{\mu_i} F_{\mu_1 \cdots \mu_J }(q)=0$. Thanks to \eqref{d=2 metric} and the transversality condition, each Minkowski dimension of tensor $ F _{\mu_1 \cdots \mu_J }(q)$ can be expanded on the basis of vectors $\partial_{ w} q^\mu,\,\partial_{\bar w} q^\mu$ and $q^\mu$, with coefficients:
\begin{align}
\label{FvecBasis}
F ^{...\mu...}(q)=\eta^{\mu\nu}F _{...\nu...}(q)&=\left(\frac{1}{2}\left(\partial_{w} q^\mu \partial_{\bar w} q^\nu+\partial_{\bar w} q^\mu\partial_{ w} q^\nu\right)+q^\mu v^\nu +v^\mu q^\nu\right)F _{...\nu...}(q)\\
&=\partial_{ w} q^\mu\left(\frac{1}{2}\partial_{\bar w} q^\nu F _{...\nu...}(q)\right)+\partial_{\bar w} q^\mu\left(\frac{1}{2}\partial_{ w} q^\nu F _{...\nu...}(q)\right)+q^\mu \left( v^\nu F _{...\nu...}(q)\right). \notag
\end{align}
However, note that in (\ref{d=2 projection}) only contractions with the terms $\partial_{ w} q^\mu$ or $\partial_{\bar w} q^\mu$ appear, and $q_\mu\partial_{ w} q^\mu=q_\mu\partial_{\bar w} q^\mu=0$. This means that only two out of three vectors of the expansion on the second line in (\ref{FvecBasis}) contribute; or, equivalently, that the coefficient of $q^\mu$ may be dropped. This tells us that while a tensor $F$ of interest in general is not expected to be transverse with respect to $v^\mu$, eq. (\ref{d=2 projection}) remains unchanged if we project $F$ to additionally be $v^\mu$-transverse $v^{\mu_i} F_{\mu_1 \cdots \mu_J }(q)=0$.\footnote{Since $v^\mu$ and $q^\mu$ are orthogonal to $\partial_{ w} q^\mu$ and $\partial_{\bar w} q^\mu$, imposing the vanishing of contributions along the directions $q^\mu$ and $v^\mu$ in tensor $F$ does not affect the projections of $F$ onto directions $\partial_{ w} q^\mu$ or $\partial_{\bar w} q^\mu$.} For convenience, we include this additional transversality in the following.

The above properties are consistent with the fact that only two tensor components on the celestial sphere are non-zero, since the mixed-index projection vanishes automatically. For example,
\begin{align}
f_{w\bar{w}}=\frac{\partial q^{\mu}}{\partial w} \frac{\partial q^{\nu}}{\partial \bar{w}} F_{\mu \nu }(q)=\left(2\eta^{\mu\nu} -\frac{\partial q^\mu}{\partial \bar{w}}\frac{\partial q^\nu}{\partial w}-2q^\mu v^\nu -2v^\mu q^\nu\right)F_{\mu \nu }(q)=-f_{\bar{w}w}=0
\end{align}
where we have used \eqref{d=2 metric} in the second step, transversality and tracelessness in the third, and symmetry of $f_{ab}$ in the fourth (a quantity that is equal to plus or minus itself must be zero). 

As before it is useful to encode $F_{\mu_1 \cdots \mu_J }(q)$ in terms of a homogeneous polynomial as
\begin{align}
F(q,Z)=F_{\mu_1 \cdots \mu_J }(q)Z^{\mu_1}\cdots Z^{\mu_J}=\mathcal F_{\mu_1 \cdots \mu_J }(q)Z^{\mu_1}\cdots Z^{\mu_J}=\mathcal F(q,Z),
\end{align}
with new auxiliary variable $Z^\mu$, which is restricted to the sub-manifold
\begin{align}\label{light cone submfd}
v^2=q^2 =q\cdot v-1=Z^2=q\cdot Z=v\cdot Z =0.
\end{align}
Here $\mathcal F_{\mu_1 \cdots \mu_J }(q)$ is $F_{\mu_1 \cdots \mu_J }(q)$ with all trace subtractions dropped due to $Z^2=0$, such that $\mathcal F_{\mu_1 \cdots \mu_J }(q)$ is symmetric transverse, but not traceless.

We present the following operator acting on the auxiliary variables $Z^\mu$ \footnote{Here the bulk dimension $\eta_{\mu\nu}\eta^{\mu\nu}=d+2$ makes its appearance, $d$ being the dimension of the celestial sphere. We keep $d$ as a parameter, since we will have to consider a limit $d\to 2$ in the following.}
\begin{align}
\label{Rop}
R_\mu =&\frac{d-2}{2}\left(\frac{\partial}{\partial Z^\mu}-v_\mu\left(q\cdot \frac{\partial}{\partial Z}\right)-q_\mu\left(v\cdot \frac{\partial}{\partial Z}\right) \right)+\left(Z\cdot \frac{\partial}{\partial Z} \right)\frac{\partial}{\partial Z^\mu}\\
&-q_\mu\left(Z\cdot \frac{\partial}{\partial Z} \right)\left(v\cdot \frac{\partial}{\partial Z} \right)
-v_\mu\left(Z\cdot \frac{\partial}{\partial Z} \right)\left(q\cdot \frac{\partial}{\partial Z} \right)\notag\\
&-\frac{Z_\mu}{2}\left(\frac{\partial^2}{\partial Z\cdot \partial Z} -2\left(v\cdot \frac{\partial}{\partial Z} \right)\left(q\cdot \frac{\partial}{\partial Z} \right)\right),\nonumber
\end{align}
which acts interior to \eqref{light cone submfd}, i.e. for any arbitrary function $g(Z)$ we have 
\begin{align}
R_\mu(v^2 g(Z))=R_\mu(q^2 g(Z))=R_\mu(Z^2 g(Z))=R_\mu(q\cdot Z g(Z))=R_\mu(v\cdot Z g(Z))=0.
\end{align}
Furthermore, $R_\mu$ is designed to be symmetric, doubly-transverse and traceless
\begin{align}
R_\mu R_\nu=R_\nu R_\mu,\qquad q^\mu R_\mu=0,\qquad v^\mu R_\mu=0, \qquad R_\mu R^\mu=0.
\end{align}
The original symmetric traceless transverse tensor $F_{\mu_1 \cdots \mu_J }(q)$ can be recovered from the auxiliary polynomial $F(q,Z)$ (or $\mathcal F(q,Z)$) by $J$-fold application of $R_\mu$ in the limit $d\to 2 :$\footnote{Note that any product of $R_\mu$ operators acting on the same number of $Z^\mu$ variables vanishes as $d\to 2$, while the Pochhammer symbol $\left(\frac{d-2}{2} \right)_J$ vanishes as well at the same rate and causes the pre-factor in (\ref{Fencode}) to diverge. Overall, the limit $d\to 2$ is finite and produces the desired symmetric traceless transverse tensor components.}
\begin{align}
\label{Fencode}
F_{\mu_1 \cdots \mu_J }(q)=\frac{1}{J!\left(\frac{d-2}{2} \right)_J}R_{\mu_1}\cdots R_{\mu_J}F(q,Z)=\frac{1}{J!\left(\frac{d-2}{2} \right)_J}R_{\mu_1}\cdots R_{\mu_J}\mathcal F(q,Z).
\end{align}
An alternative expression for $R_\mu$ is obtained by making use of \eqref{d=2 metric} in \eqref{Rop}, so that
\begin{align}
R_\mu =&\frac{d-2}{4}\left((\partial_{w}q_\mu)\left((\partial_{\bar w}q)\cdot \frac{\partial}{\partial Z}\right)+(\partial_{\bar w}q_\mu)\left((\partial_{w}q)\cdot \frac{\partial}{\partial Z}\right) \right)-\frac{Z_\mu}{2}\left((\partial_{\bar w}q)\cdot \frac{\partial}{\partial Z} \right)\left((\partial_{ w}q)\cdot \frac{\partial}{\partial Z} \right)\notag\\
&+\frac{1}{2}(\partial_{\bar w}q_\mu)\left(Z\cdot \frac{\partial}{\partial Z} \right)\left((\partial_{w}q)\cdot \frac{\partial}{\partial Z} \right)
+\frac{1}{2}(\partial_{ w}q_\mu)\left(Z\cdot \frac{\partial}{\partial Z} \right)\left((\partial_{\bar w}q)\cdot \frac{\partial}{\partial Z} \right).
\end{align}
Due to
\begin{align}
\label{partialQcontr}
\partial_{w} q^{\mu}\partial_{w} q^{\nu}\eta_{\mu\nu}=\partial_{\bar w} q^{\mu}\partial_{\bar w} q^{\nu}\eta_{\mu\nu}=0\qquad\text{and}\qquad \partial_{\bar w} q^{\mu}\partial_{ w} q^{\nu}\eta_{\mu\nu}=2\,,
\end{align}
we have the simplifications
\begin{align}
\label{DqR1}
(\partial_{w}q)\cdot R =&\frac{d-2}{2}(\partial_{w}q)\cdot \frac{\partial}{\partial Z} -\frac{(\partial_{w}q)\cdot Z}{2}\left((\partial_{\bar w}q)\cdot \frac{\partial}{\partial Z} \right)\left((\partial_{ w}q)\cdot \frac{\partial}{\partial Z} \right)+\left(Z\cdot \frac{\partial}{\partial Z} \right)\left((\partial_{ w}q)\cdot \frac{\partial}{\partial Z} \right),\notag\\
(\partial_{\bar w}q)\cdot R =&\frac{d-2}{2}(\partial_{\bar w}q)\cdot \frac{\partial}{\partial Z} -\frac{(\partial_{\bar w}q)\cdot Z}{2}\left((\partial_{\bar w}q)\cdot \frac{\partial}{\partial Z} \right)\left((\partial_{ w}q)\cdot \frac{\partial}{\partial Z} \right)+\left(Z\cdot \frac{\partial}{\partial Z} \right)\left((\partial_{\bar w}q)\cdot \frac{\partial}{\partial Z} \right).
\end{align}
Applying (\ref{partialQcontr}) repeatedly in the powers $\left((\partial_{w}q)\cdot R\right)^J$ and $\left((\partial_{\bar w}q)\cdot R\right)^J$, we observe
\begin{align}
\label{DqR2}
f_J(w,\bar{w})&=\frac{\left((\partial_{w}q)\cdot R\right)^J}{J!\left(\frac{d-2}{2} \right)_J}\mathcal F(q,Z)=\frac{1}{J!}\left((\partial_{w}q)\cdot \frac{\partial}{\partial Z}\right)^J \mathcal F(q,Z) = (\partial_{w}q^{\mu_1})...(\partial_{w}q^{\mu_J})\mathcal F_{\mu_1 \cdots \mu_J }(q),\notag\\
\tilde f_J(w,\bar{w})&=\frac{\left((\partial_{\bar w}q)\cdot R\right)^J}{J!\left(\frac{d-2}{2} \right)_J}\mathcal F(q,Z)=\frac{1}{J!}\left((\partial_{\bar w}q)\cdot \frac{\partial}{\partial Z}\right)^J \mathcal F(q,Z) = (\partial_{\bar w}q^{\mu_1})...(\partial_{\bar w}q^{\mu_J})\mathcal F_{\mu_1 \cdots \mu_J }(q),
\end{align}
which means that all spin plus or all spin minus pull-back of a symmetric traceless transverse tensor $F$ to the celestial sphere is equivalent to the same pull-back of the corresponding symmetric transverse (not-traceless) tensor $\mathcal F$, which explains the simplification in (\ref{GtransSym}).

For two spin-$J$ symmetric traceless tensors $f_1$ and $f_2$ on $\mathbb{C}$ the tensor contraction is given by
\begin{align}
g^{a_1 b_1}\cdots g^{a_J b_J} f_1{}_{a_1 \cdots a_J} f_2{}_{b_1 \cdots b_J}= \frac{f_1{}_J \tilde{f}_2{}_J + \tilde{f}_1{}_J f_2{}_J}{2^J}.
\end{align}
Using \eqref{d=2 metric}  and the fact that $F_1$ and $F_2$ are transverse and traceless, it is easy to see that this is equal to the contraction between their embedding space counterparts \cite{Costa:2011mg}
\begin{align}
  \frac{f_1{}_J \tilde{f}_2{}_J + \tilde{f}_1{}_J f_2{}_J}{2^J}=F_1^{\mu_1 \cdots \mu_J}F_2{}_{\mu_1 \cdots \mu_J}=\frac{1}{J! \left( \frac{d-2}{2}\right)_J}\mathcal F_1(q,R)\mathcal F_2(q,Z).
\end{align}
Since both, $\mathcal F_1$ and $\mathcal F_2$ are transverse, all occurrences of $q^\nu$ in the definition (\ref{Rop}) of operator $R_\mu$ vanish, so that $R_\mu$ simplifies to an operator called $D_{Z^\mu}$ and the contraction effectively reads
\begin{align}
\frac{1}{J! \left( \frac{d-2}{2}\right)_J}\mathcal F_1(q,R)\mathcal F_2(q,Z)=\frac{1}{J! \left( \frac{ d-2}{2}\right)_J}\mathcal F_1(q,D_Z)\mathcal F_2(q,Z).
\end{align}
The simpler operator $D_{Z^\mu}$ is given by
\begin{align}\label{trace op d=2}
D_{Z^\mu} =\left(\frac{ d}{2}-1+Z\cdot \frac{\partial}{\partial Z} \right)\frac{\partial}{\partial Z^\mu}-\frac{1}{2}Z_\mu \frac{\partial^2}{\partial Z\cdot \partial Z}.
\end{align}
This explicitly demonstrates the connection with eqs. (3.29) and (3.30) in \cite{Costa:2011mg}.

\section{Example spinning massive amplitudes mapped to the celestial sphere}
\label{sec:3pexamples}
The simplest examples of amplitudes involving at least one massive spinning external particle are three-point amplitudes of a massive spinning particle of spin $s$ and two massless scalars. In Minkowski space such three-point couplings are fixed to 
\begin{align}
\mathcal{A}_3{}_a\sim {\epsilon_1}_a^{\mu_1...\mu_s}(p_2-p_3)_{\mu_1}...(p_2-p_3)_{\mu_s}\delta^{(4)}(p_1^\mu-p_2^\mu-p_3^\mu),
\end{align}
where the massive polarization tensor ${\epsilon_1}_a^{\mu_1...\mu_s}$ for the first particle, e.g., in case of spin $s=1$ or $s=2$ is parametrized as in (\ref{s1pols}) or (\ref{s2pols}), and momenta $p^\mu_1=m \hat p^\mu_1$ and $p_2^\mu=\omega_2 q_2^\mu,~p_3^\mu=\omega_3 q_3^\mu$ as in (\ref{mphat}) and (\ref{omegaq}).

According to the mapping prescriptions discussed around (\ref{massAmap}), this amplitude is transformed to the celestial sphere as
\begin{align}
\label{example3pt}
A{}^{(s)}_{3\,J}=\int_0^\infty\frac{dy_1}{y_1^3}\int dz_1 d\bar z_1 \sum_a {G_1}^{(s)}_{Ja}\left(\prod_{i=2}^3\int_0^\infty d\omega_i\, \omega_i^{\Delta_i-1}\right) \mathcal{A}_3{}_a.
\end{align}
For any spin $s$ of the massive particle, the momentum conservation delta functions are localized by the integrals in the same way, the solution being
\begin{align}
\label{sol3}
y_1=&\frac{m}{2(\omega_2+\omega_3)}~~~,~~~z_1=\frac{w_2\omega_2+w_3\omega_3}{\omega_2+\omega_3}~~~,~~~\bar z_1=\frac{\bar w_2\omega_2+\bar w_3\omega_3}{\omega_2+\omega_3}\,,
\end{align}
and
\begin{align}
\label{sol3w}
\omega_2=\frac{m^2}{4(w_2-w_3)(\bar w_2-\bar w_3)\omega_3},
\end{align}
with absolute value of delta function integration Jacobian given by
\begin{align}
|J|=\frac{8m(w_2-w_3)^2(\bar w_2-\bar w_3)^2\omega_3^2}{(m^2+4(w_2-w_3)(\bar w_2-\bar w_3)\omega_3^2)^3}.
\end{align}
Having saturated four out of five integrations by localizing the momentum conservation delta functions, the remaining $\omega_3$ integration in all $J$ components can be performed by making use of the integral
\begin{align}
\int_0^\infty d\omega\, \omega ^c \left(a+b\, \omega ^2\right)^d =\frac{1 }{2 }b^{-\frac{c}{2}-\frac{1}{2}} 
   a^{\frac{c}{2}+d+\frac{1}{2}} {\mathcal{B}} \left(\frac{c+1}{2},-\frac{c}{2}-d-\frac{1}{2}\right)\,,
\end{align}
which holds for Re$(c)>-1$, Re$(a)>0$, Re$(b)>0$, Re$(c+2d)<-1$, and where $\mathcal{B}(x,y)$ is the Euler beta function.

\subsection{Celestial amplitude of two massless scalars and one massive spin-1 boson}
In case when the spin of the massive particle is $s=1$, the above calculation (\ref{example3pt}) leads to
\begin{align}
A^{(1)}_{3\,J}=C^{J}_{\Delta_1,\Delta_2,\Delta_3}
w_{12}^{\frac{-\Delta _1-\Delta _2+\Delta
   _3-J}{2} }  w_{23}^{\frac{\Delta _1-\Delta _2-\Delta _3+J}{2}
   }w_{31}^{\frac{-\Delta
   _1+\Delta _2-\Delta _3-J}{2} }
   \bar{w}_{12}^{\frac{-\Delta _1-\Delta
   _2+\Delta _3+J}{2} } 
   \bar{w}_{23}^{\frac{\Delta _1-\Delta
   _2-\Delta _3-J}{2} }\bar{w}_{31}^{\frac{-\Delta _1+\Delta _2-\Delta _3+J}{2}
   },
\end{align}
with $w_{i,j}=w_i-w_j$, $\bar w_{i,j}=\bar w_i-\bar w_j$ and $J=-1,0,1$. The resulting $J$ components of the three-point function coefficient can be summarized as
\begin{align}
C^{J}_{\Delta_1,\Delta_2,\Delta_3}= \frac{J^2-2 J-1}{2^2} \left(\frac{m}{2}\right)^{\Delta
   _2+\Delta _3-3} \left(\frac{\Delta _2-\Delta
   _3}{\Delta _1}\right)^{1-\left| J\right| }
   \mathcal{B}\left(\frac{\left| J\right| +\Delta
   _1+\Delta _2-\Delta _3}{2} ,\frac{\left|
   J\right| +\Delta _1-\Delta _2+\Delta _3}{2} \right).
\end{align}
This matches the multiplet constructed from the highest weight state (\ref{C3m00ans}) when $c=const.$.

\subsection{Celestial amplitude of two massless scalars and one massive spin-2 boson}
In case when the spin of the massive particle is $s=2$, the above calculation (\ref{example3pt}) leads to
\begin{align}
A^{(2)}_{3\,J}=C^{J}_{\Delta_1,\Delta_2,\Delta_3}
w_{12}^{\frac{-\Delta _1-\Delta _2+\Delta
   _3-J}{2} }  w_{23}^{\frac{\Delta _1-\Delta _2-\Delta _3+J}{2}
   }w_{31}^{\frac{-\Delta
   _1+\Delta _2-\Delta _3-J}{2} }
   \bar{w}_{12}^{\frac{-\Delta _1-\Delta
   _2+\Delta _3+J}{2} } 
   \bar{w}_{23}^{\frac{\Delta _1-\Delta
   _2-\Delta _3-J}{2} }\bar{w}_{31}^{\frac{-\Delta _1+\Delta _2-\Delta _3+J}{2}
   },
\end{align}
with $w_{i,j}=w_i-w_j$, $\bar w_{i,j}=\bar w_i-\bar w_j$ and $J=-2,-1,0,1,2$. The resulting $J$ components of the three-point function coefficient can be summarized as
\begin{align}
C^{J}_{\Delta_1,\Delta_2,\Delta_3}=& \frac{ \left(\frac{m}{2}\right)^{\Delta _2+\Delta
   _3-2} 
   \left( \left(2-\Delta _1\right) \Delta _1
   \delta _{0,J}-\frac{1}{4} (J (J ((8-7 J)
   J+19)-32)-12) \left(\Delta _2-\Delta
   _3\right){}^{2-\left| J\right| }\right)}{6\left(\left|
   J\right| +\Delta _1\right)_{2-\left| J\right| }}\notag\\
	&\mathcal{B}\left(\frac{1}{2} \left(\left| J\right| +\Delta
   _1+\Delta _2-\Delta _3\right),\frac{1}{2} \left(\left|
   J\right| +\Delta _1-\Delta _2+\Delta _3\right)\right),
\end{align}
where  $(x)_y={\Gamma(x+y)}/{\Gamma(x)}$ is the Pochhammer symbol. Again, this matches the multiplet constructed from the highest weight state (\ref{C3m00ans}) when $c=const.$.

%
%This is the most common positions for acknowledgments. A macro is
%available to maintain the same layout and spelling of the heading.
%
%\paragraph{Note added.} This is also a good position for notes added
%after the paper has been written.
% BIBLIOGRAPHY
% use BIBTEX if you want
%\bibliographystyle{JHEP}
%\bibliography{yourBIBfiles}

% The bibliography will probably be heavily edited during typesetting.
% We'll parse it and, using the arxiv number or the journal data, will
% query inspire, trying to verify the data (this will probalby spot
% eventual typos) and retrive the document DOI and eventual errata.
% We however suggest to always provide author, title and journal data:
% in short all the informations that clearly identify a document.

%

\end{document}